\documentclass{article}

\usepackage[english]{babel}

\usepackage[letterpaper,top=2cm,bottom=2cm,left=3cm,right=3cm,marginparwidth=1.75cm]{geometry}

\usepackage{amsmath}
\usepackage{graphicx}
\usepackage[colorlinks=true, allcolors=blue]{hyperref}
\usepackage{caption}
\usepackage{subcaption}
\usepackage{amsthm}
\usepackage{comment}
\usepackage{amsthm}
\usepackage{algorithm}
\usepackage{bm}
\usepackage{multirow}
\usepackage{array}
 \usepackage{booktabs}
\usepackage[noend]{algpseudocode}
\newcommand\numberthis{\addtocounter{equation}{1}\tag{\theequation}}
\newcommand\independent{\protect\mathpalette{\protect\independenT}{\perp}}
\def\independenT#1#2{\mathrel{\rlap{$#1#2$}\mkern2mu{#1#2}}}

\usepackage{tikz}
\usetikzlibrary{fit}
\usetikzlibrary{arrows.meta,positioning}
\usetikzlibrary{calc}

\newtheorem{theorem}{Theorem}
\newtheorem{definition}{Definition}

\newtheorem{remark}{Remark}
\newtheorem{corollary}{Corollary}
\newtheorem{proposition}{Proposition}
\newtheorem{assumption}{Assumption}

\allowdisplaybreaks
\newcommand{\given}{\,|\,}

\usepackage{xspace}
\usepackage{bbm}
\usepackage{mathrsfs}
\usepackage{amssymb}

\newcommand{\Ec}{\mathcal{E}}

\newcommand{\Hc}{\mathcal{H}}

\newcommand{\Lc}{\mathcal{L}}
\newcommand{\Nc}{\mathcal{N}}

\newcommand{\Sc}{\mathcal{S}}

\newcommand{\Wc}{\mathcal{W}}
\newcommand{\Xc}{\mathcal{X}}
\newcommand{\Yc}{\mathcal{Y}}

\newcommand{\Xv}{{\bf X}}
\newcommand{\Yv}{{\bf Y}}

\def\d{\delta}

\DeclareMathOperator\E{\sf E}
\let\P\relax
\DeclareMathOperator\P{\sf P}
\DeclareMathOperator\R{\sf R}

\def\textiid{i.i.d.\@\xspace}
\newcommand\iid{\ifmmode\text{ i.i.d. } \else \textiid \fi}

\newcommand{\Real}{\mathbb{R}}

\title{Partial Invariance for Classification under \\Structural Causal Models}
\author{Austin Goddard$^{\displaystyle 1}$ and Yu Xiang$^{\displaystyle 2}$\\
$^{\displaystyle 1}$Department of Electrical and Computer Engineering, University of Utah\\
$^{\displaystyle 2}$Department of Mathematics and Statistics, Florida Atlantic University}

\begin{document}
\maketitle

\begin{abstract}
    Making predictions in an unseen environment from data collected across multiple training environments is challenging. This challenge is particularly pronounced when the target-generating mechanism varies across environments. In this work, we introduce \emph{environment-label separation}, a form of partial environmental invariance that decomposes the target conditional distribution into an environment-invariant component and a label-invariant component. We show that this partial invariance enables estimation of the target conditional distribution in an unseen environment when using labeled training data together with unlabeled test features. Within the structural causal model framework, we provide sufficient conditions for environment-label separation. Building on these conditions, we develop an algorithm, dIMP-Classifier, which jointly minimizes an estimate of the test risk and encourages invariance across environments. We further provide finite-sample guarantees that characterize the test risk in terms of deviations from invariance, estimation error, and sample complexity. Experiments on simulated data and the real-world light tunnel Causal Chamber data demonstrate that the proposed approach can perform under interventions on the target mechanism and consistently outperforms the considered standard and invariance-based baselines.
\end{abstract}

\section{Introduction and Problem Formulation}
It is common practice to collect observations of a set of features and response from different environments to train a model. In this setting, the joint distribution of the features and the response may vary from environment to environment, particularly from the training environment to the test environment. Such distribution shifts pose a fundamental challenge to generalization, as statistical relationships learned in the training environments need not persist in the test environment. Consider the classification of a discrete random variable $Y^e$ given the set of features $X^e=\{X_1^e,\dots, X_m^e\}$ for different environmental conditions indexed by the set $\Ec:= \Ec_{\text{train}}\cup\{e^{\text{test}}\}$, which consists of all the training environments $\Ec_{\text{train}}$ and one test environment $e^{\text{test}}$. In the test environment, let $(X^t,Y^t):=(X^{e^{\text{test}}},Y^{e^{\text{test}}})$. While we observe the variables $(X^e,Y^e)$ for all $e\in\Ec_{\text{train}}$, only the variable $X^t$ is observed in the test environment. The goal is to make predictions on $Y^t$ given $\{(X^e,Y^e): e\in\Ec_{\text{train}}\} \cup \{X^t\}$. We assume that the supports of $X^e$ and $Y^e$ do not vary with environment and are $\Xc$ and $\Yc$, respectively.

One principled way to address this problem is to model distribution shifts as changes in the underlying data-generating process using structural causal models (SCMs)~(\cite{pearl2009causality,peters2017elements}). This perspective makes it possible to characterize which mechanisms remain stable across environments and which invariant relationships may therefore be exploited for prediction. The pioneer work of invariant causal prediction (ICP)~\cite{peters2016causal} initiated a rich line of results (e.g.,~\cite{heinze2018invariant,pfister2019invariant,runge2023causal,toth2022active}) following the \emph{invariance principle} (also see~\cite{haavelmo1944probability,aldrich1989autonomy,pearl2009causality,imbens2015causal}), which requires that $\P(Y^e|X_S^e)$ be invariant with respect to $e$, where $X^e_{S}$ denotes a subset of the random variables in $X^e$. In practical settings, requiring the entire conditional distribution of the target to remain invariant across environments can be overly restrictive, particularly when the target mechanism is subject to change. This motivates the use of weaker forms of invariance that preserve only certain components or relationships across environments while allowing the target conditional distribution itself to vary.


In this work, we introduce a form of \emph{partial} environmental invariance for \emph{classification} tasks where the label $Y^e$ is discrete and takes on values from $[1:L]:=\{1,...,L\}$. We wish to identify a classifier $h:\Xc_S\rightarrow \Yc \in\Hc$, for some $S\subseteq [1:m]$, such that the target risk is small, where the risk for a given environment $e\in\Ec$ is defined as
\begin{equation}\label{equ::risk}
    \R_S^e(h) := \E[\mathbf{1}_{h(X_S^e)\ne Y^e}] = \P(h(X_S^e)\ne Y^e). 
\end{equation}
The target risk (i.e., the risk for environment $e^{\text{test}}$) of a classifier $h$ is denoted by $\R_S^t(h)$. 
Without further assumptions on the relationship between environments, there is no guarantee that the risk of any classifier constructed using training data from various environments will be close to that of $h^{\text{oracle}} := \arg\min_{h\in\Hc} \, \R_S^t(h)$. In an attempt to remedy this, our objective is to identify consistent relationships between all environments such that we can successfully minimize the risk over the test environment. For this to be possible, we consider a form of \emph{partial} environmental invariance in which the probability of $Y^e$ given $X_S^e$ can be separated into environmentally variant and invariant components. We call it \emph{environment-label separation} or \emph{EY-separation} for short. 


\begin{definition}[EY-Separation]\label{def::parial_invar}
For all $(x,y)\in\Xc\times\Yc$ and $e\in \Ec$, if there exist $S\subseteq [1:m]$, $\psi^e:\Xc_S\rightarrow \Wc$, and $\rho_y: \Xc_S\times\Wc \rightarrow [0,1]$, such that $\P(Y^{e}=y\given X_S^e=x)$ admits the following factorization,
\begin{equation}\label{equ::partial_invar}
        \P(Y^{e}=y\given X_S^e=x) = \underbrace{\rho_y}_{\text{invariant to $e$}}\bigl(x,\underbrace{\psi^e(x)}_{\text{varying with $e$}}\bigr),
    \end{equation}
     we say that $\P(Y^{e}=y\given X^e_S=x)$ is EY-separable\footnote{We say that $\P(Y^{e}=y\given X^e_S=x)$ is EY-separable and that $\rho_y$ and $\psi^e$ are EY-separable functions interchangeably, both referring to the factorization in Definition~\ref{def::parial_invar}.}. 

\end{definition}

 Note that $\rho_y$ is not dependent on $e$ and $\psi^e$ is not dependent on $y$. We stress that only $\psi^e$ is allowed to change with environments, and it should not depend on $y$ since the label is not available in the test environment. This enables flexible domain adaptation in the test environment $e=e^{\text{test}}$, because $\psi^t:= \psi^{e^{\text{test}}}$ does not rely on the test label $Y^{t}$, making its estimation possible at test time (we will elaborate on this crucial point shortly in~\eqref{eq::EY_estimate}). We thus only require \emph{partial invariance} via $\rho_y$, while keeping $\psi^e$ adaptive. This is in contrast to the existing works (e.g.,~\cite{peters2016causal}) that require invariance for the whole conditional distribution. 

 \begin{remark}
     We note that the set $S$ that satisfies~\eqref{equ::partial_invar} is determined by the joint distribution of $(X,Y)$. In general, the set $S$ is not unique, meaning that various $S$ can induce representations that permit EY-separation.  We give examples of such settings in  Section~\ref{sec::mot_examp} and present specific sufficient conditions for $S$ in Condition~$2$ of Proposition~\ref{prop::suff2}.   
 \end{remark}

\smallskip
\noindent{\bf How to make use of EY-separation?} The identified EY-separable functions can lead to efficient classification in the test environment: 
\begin{enumerate}
    \item As $\rho_y$ does not depend on $e$, an estimate of $\rho_y$ trained on any (or all) training environments will extend to the test environment. 
    \item As $\psi^e$ does not depend on $y$, it is possible to estimate using \emph{only} the test features, even though labels for the test environment are not available. 
\end{enumerate}
As a result, identifying EY-separable functions gives a path to estimate $\P(Y^t|X^t_S)$ without access to test labels. For each class label $y\in [1: L]$,  we can estimate $P(Y^t=y \,|\, X^t_S=x)$ in the test environment $ e^{\text{test}}$ using only the test features $X_S^{t}$ as 
\begin{equation}
\label{eq::EY_estimate}
        \widehat{\P(Y^t=y \,|\, X^t_S=x)} = \underbrace{\widehat{\rho_{\,y}}}_{\text{learned from $\Ec_{\text{train}}$}}\bigl(\,x,\, \underbrace{\widehat{\psi^{t}(x)}}_{\text{plug in test features }  X_S^{t}}\,\bigr), 
    \end{equation}
    which then leads to a natural classifier that chooses the label with the largest conditional probability. 


\medskip
Our main contributions are summarized as follows. We introduce a form of partial environmental invariance for classification called environment-label separation and show that it allows for estimation of the target without access to test labels. We provide sufficient conditions under the SCM framework that guarantee environment-label separation (Section~\ref{sec:: EY_SCM}). We then develop practical prediction and identification procedures using separable functions (Section~\ref{sec::methods}) and provide finite-sample analysis of its performance (Section~\ref{sec::ident_CP}). Finally, in Section~\ref{sec::experiments}, we evaluate the proposed approach on both simulated data and the real-world light tunnel Causal Chamber benchmark~\cite{gamella2025causal}, a physically grounded optical system in which controlled interventions induce distribution shifts across environments. We show that the proposed method consistently outperforms the considered baselines, including existing invariant prediction approaches.






\subsection{Test Risk via EY-separation}
While identifying EY-separable functions gives a path to estimate $\P(Y^t|X_S^t)$ without access to test labels, it does not guarantee that the proposed estimate will perform well at test time (i.e., small test risk). Thus, we explore an important property of EY-separation. Consider any classifier $h(x_S)$, if $\rho_y$ and $\psi^t$ are EY-separable according to~\eqref{equ::partial_invar}, we introduce the following test risk.

\smallskip
\begin{definition}[Risk with EY-Separation]
    When EY-separation holds, we define the corresponding risk $\R^t_{\text{EY},S}$ as 
\begin{equation}
    \R^t_{\text{EY},S}(h):= \E[1-\rho_{h(X^t_S)}(X^t_S,\psi^t(X^t_S))].  
\end{equation}
\end{definition}

\smallskip
We make the following simple observation under EY-separation, 
\begin{align}\label{equ::EY_to_R}
\R_S^t(h) = \R^t_{\text{EY},S}(h), 
\end{align} 
which follows from $\P(Y^t \ne h(X_S^t)) 
= \E[\P(Y^t \ne h(X_S^t)\,|\,X_S^t)]
= \E[1-\rho_{h(X^t_S)}(X^t_S,\psi^t(X^t_S))]$. When the EY-separation does not hold, we can still define a general target risk for generic functions $\widetilde{\rho}$ and $\widetilde{\psi}$ 
 \begin{equation}\label{equ::EY_risk}
    \widetilde{\R}_S^t(h) := \widetilde{\R}_S^t(h\;;\;\boldsymbol{\widetilde{\rho}},\widetilde{\psi}):= \E[1-\widetilde{\rho}_{h(X^t_S)}(X_S^t,\;\widetilde{\psi}(X_S^t) )],
\end{equation}
 where $\widetilde{\rho}_l$ depends on $l\in [1:L]$ and can be dependent on $t$ as well, and $\boldsymbol{\widetilde{\rho}}:= (\widetilde{\rho}_l)_{l\in[1:L]}$, while $\widetilde{\psi}$ depends on $t$ but not on $y$ (again since the label is not available at test time). $\boldsymbol{\widetilde{\rho}}$ and  $\widetilde{\psi}$ should be viewed as dummy variables that will be specified in different ways as needed in our analysis in Section~\ref{sec::methods}.

\smallskip
Assuming  $\rho_y$ and $\psi^e$ are EY-separable,~\eqref{equ::EY_to_R} shows that minimizing the target EY-separation risk will in turn minimize the risk over the target environment. Thus, with the ability to estimate $\P(Y^t|X_S^t)$ without test labels, and the ability to minimize test risk according to~\eqref{equ::EY_to_R}, our roadmap moving forward is to: (I) determine under what conditions EY-separation is satisfied, and (II) create a practical methodology that leverages the useful properties of EY-separation.

\subsection{Structural Causal Models (SCMs) and Notation}
Our work is developed based on the structural causal models (SCMs)~\cite{pearl2009causality,peters2017elements} and we adopt the conventional notation (e.g., see~\cite{lauritzen2001causal}). That is, the predictors $\text{pa}(X^e_i)\subseteq X^e\cup Y^e \backslash X_i^e$ and $\text{pa}(Y^e) \subseteq X^e$ denote the set of direct causal parents of $X^e_i$ and $Y^e$, respectively. Similarly, we refer to $X_i^e$ as a causal child of some $X^e_j \in \text{pa}(X^e_i)$. Additionally, we denote $de(X^e_i)$, $nd(X^e_i)$, and $an(X^e_i)$ as descendants, nondescendants, ancestors of $X^e_i$, respectively. For some $X^e_i$ and $X^e_j$, we often denote $X^e_i$ being a parent of $X^e_j$ as $X^e_i\rightarrow X^e_j$. The variables $X^e$ and $Y^e$ are part of an \emph{acyclic} SCM $\Sc^e$
\begin{equation} \label{equ::SCM}
    \Sc^e: \begin{cases}
    Y^e\leftarrow f_Y^e(\text{pa}(Y^e)\, ,\ \epsilon^e_Y),\\
    X_1^e \leftarrow f_1^e(\text{pa}(X_1^e)\, ,\ \epsilon^e_1),\\
    \qquad \vdots \\
    X_m^e \leftarrow f_m^e(\text{pa}(X_m^e)\, ,\ \epsilon^e_m),	\end{cases}
\end{equation}
where $\epsilon^e_Y$, and $\epsilon_0^e,\dots,\epsilon_m^e$ are \emph{independent} noise variables with distribution $\epsilon_i^e\sim \mathsf{F}_i^e$, and the graph induced by $\Sc^e$ is acyclic, meaning that a predictor $X_i^e$ cannot be both a descendant and an ancestor of another variable $X_j^e$. We note that the underlying graph structure of $\Sc^e$ does not change over environment as the function $\text{pa}(\cdot)$ does not change over environment. 


An intervention on a variable occurs if the structural assignment or noise distribution changes for some $e\in\Ec$. Many approaches~\cite{peters2016causal,pfister2021stabilizing,du2023learning} make various assumptions on the types of interventions (e.g., shifts in the mean or variance) and which variables can and cannot be intervened. This formulation allows general interventions on most variables (see Proposition~\ref{prop::suff} for the few exceptions) and on the variable $Y^e$, which many other approaches fail to address.

\subsection{Related Works}\label{sec::related_works}

The authors of~\cite{chen2021domain} provided a systematic treatment of domain adaptation using SCMs, leading to the conditional invariant residual matching (CIRM) method. The CIRM and its variants leverage the domain invariant projection (DIP)-type methods~(also see \cite{baktashmotlagh2013unsupervised,ganin2016domain} and the generalized label shift to handle target label perturbation~\cite{li2019target,tachet2020domain}). Theoretical guarantees have been provided for the prediction performance under shift interventions on $Y$, while numerical studies are provided for interventions on the noise variance of $Y$~\cite{chen2021domain}.  More recently, the role of conditional invariance penalty~\cite{gong2016domain,heinze2021conditional}, which assumes the existence of conditional invariant components (CICs) in the anticausal setting, has been highlighted. It has also been pointed out that the general mixed-causal-anticausal domain adaptation problem remains open. To the best of our knowledge, all the existing frameworks, that concern either $\P(\phi(X^e)\,|\,Y^e)$ (e.g., \cite{gong2016domain,heinze2021conditional,
wu2025prominent}) or $\P(Y^e\,|\,\phi(X^e))$ (e.g., \cite{arjovsky2019invariant}), only assume the \emph{existence} of transforms without constructions and the transforms are \emph{not} allowed to vary with $e$. To construct the invariant transform for linear SCMs, the authors proposed a systematic way through the invariant matching property (IMP)~\cite{du2023learning,du2023generalized}, when the conventional invariance principle fails, namely when the target variable is under intervention, with an \emph{explicit construction} of transform  (also see real-world applications in~\cite{du2023learning,du2023identifying,goddard2024multi}). In this work, our goal is to extend this to the nonlinear setting, where $Y$ is discrete/categorical in classification problems; it turns out that this is more challenging and requires a different set of tools.

Perhaps the most closely related work is the sparse joint shift (SJS) framework introduced in~\cite{chen2022estimating} (also see~\cite{tasche2023sparse}). SJS permits distribution shifts in a subset of features $X^e_I$, while assuming that the conditional distribution of remaining features $X^e_S$ is invariant given $X^e_I$ and $Y^e$. This condition is closely related to the dIMP property introduced in~\eqref{equ::dimp_def} that states for subsets of variables $X_S$ and $X_C$, $\E[X_C^e\given X_S^e=x,Y^e=y]$ is invariant with respect to environments. Despite being closely related, there are important differences between the two. First, SJS is a stronger condition compared to the dIMP as SJS assumes invariance over a conditional distribution while the dIMP requires only invariance over the conditional expectation. Another important difference is the underlying problem being solved: while we consider a multinomial classification problem where we construct a classifier to estimate $\P(Y^t\given X^t_S)$, the methods proposed in~\cite{chen2022estimating} already assume the existence of a classifier and mean to estimate the performance of said classifier on a test environment. The methods developed in~\cite{chen2022estimating} are based on the reweighting scheme and thus do not provide methods for estimating $\P(Y^t\given X^t_S)$ in the test environment. Finally, the invariance exploited in the dIMP is derived from sufficient conditions, building on the SCMs in~\eqref{equ::SCM}, which helps provide specific causal interpretation to the invariant quantities that~\cite{chen2022estimating,tasche2023sparse} do not consider. 

The role of causality in facilitating domain adaptation problems is first articulated in~\cite{scholkopf2012causal}, focusing on causal and anticausal predictions. Reweighting methods have been extensively studied for covariate shift~\cite{quinonero2008dataset,storkey2009training,sugiyama2012machine}, which assumes that only the feature distribution changes over environments while the conditionals remain the same. The label shift, which aligned with the anticausal setting, has attracted much attention recently~\cite{lipton2018detecting,azizzadenesheli2018regularized,garg2020unified}. Many other interesting domain adaptation methods have been developed but they are less related to this work. The performance bounds using Vapnik-Chervonenkis (VC) theory have been initiated in~\cite{ben2006analysis}. There are fundamental works from the robust statistics perspective including distributional robust learning~\cite{bagnell2005robust,hu2013kullback,sinha2018certifying,gao2017distributional,lee2018minimax,duchi2021learning} and adversarial machine learning~\cite{goodfellow2018making,raghunathan2018semidefinite}.




\section{How to Construct EY-Separation? A Motivating Example}\label{sec::mot_examp}
\begin{figure}[h]
\centering

\begin{subfigure}[t]{0.48\textwidth}
\centering
\resizebox{0.88\linewidth}{!}{%
\begin{tikzpicture}[
    >=Stealth,
    every node/.style={font=\footnotesize},
    var/.style={
      circle,
      draw,
      minimum size=1.0cm,
      inner sep=1pt,
      align=center
    },
    target/.style={
      circle,
      draw,
      very thick,
      minimum size=1.1cm,
      inner sep=1pt,
      align=center
    },
    faint/.style={
      circle,
      draw=gray!45,
      text=gray!55,
      minimum size=1.0cm,
      inner sep=1pt,
      align=center
    },
    region/.style={
      draw,
      dashed,
      rounded corners,
      inner sep=7pt
    }
  ]

\node[var]    (H)  at (-1.5,1.8)    {$X_0$};
\node[var]    (X1) at (1.5,1.8)  {$X_{1}$};
\node[target] (Y)  at (0,0)      {$\mathbf{Y}$};
\node[var]    (X2) at (0,-1.8)   {$X_{2}$};

\node[faint] (Z1) at (-3.4, 1)    {$X_{5}$};
\node[faint] (Z3) at (-3.4,-1.8)  {$X_{4}$};
\node[faint] (Z4) at (3.4,-1.8)   {$X_{3}$};

\draw[->, dashed, gray!90] (Y) -- (Z3);
\draw[->, dashed, gray!90] (Y) -- (Z4);
\draw[->, dashed, gray!90] (Z1) -- (Z3);
\draw[->, dashed, gray!90] (H) -- (Y);
\draw[->, dashed, gray!90] (X1) -- (Y);

\draw[->] (Y) -- (X2);
\draw[->] (X1) -- (X2);

\node[region, fit=(X1)(Y)(X2)] (R) {};

\end{tikzpicture}

}
\caption{}
\label{fig:motivating_example_DAG_a}
\end{subfigure}
\hfill
\begin{subfigure}[t]{0.48\textwidth}
\centering
\resizebox{0.88\linewidth}{!}{%
\begin{tikzpicture}[
    >=Stealth,
    every node/.style={font=\footnotesize},
    var/.style={
      circle,
      draw,
      minimum size=1.0cm,
      inner sep=1pt,
      align=center
    },
    target/.style={
      circle,
      draw,
      very thick,
      minimum size=1.1cm,
      inner sep=1pt,
      align=center
    },
    faint/.style={
      circle,
      draw=gray!45,
      text=gray!55,
      minimum size=1.0cm,
      inner sep=1pt,
      align=center
    },
    region/.style={
      draw,
      dashed,
      rounded corners,
      inner sep=7pt
    }
  ]

\node[var]    (H)  at (-1.5,1.8)    {$H$};
\node[var]    (X1) at (1.5,1.8)  {$X_{1}$};
\node[target] (Y)  at (0,0)      {$\mathbf{Y}$};
\node[var]    (X2) at (0,-1.8)   {$X_{2}$};

\node[faint] (Z1) at (-3.4, 1)    {$X_{5}$};
\node[faint] (Z3) at (-3.4,-1.8)  {$X_{4}$};
\node[faint] (Z4) at (3.4,-1.8)   {$X_{3}$};

\draw[->, dashed, gray!90] (Y) -- (Z3);
\draw[->, dashed, gray!90] (Y) -- (Z4);
\draw[->, dashed, gray!90] (Z1) -- (Z3);
\draw[->, dashed, gray!90] (H) -- (Y);
\draw[->, dashed, gray!90] (H) -- (X1);
\draw[->, dashed, gray!90] (X1) -- (Y);

\draw[->] (Y) -- (X2);
\draw[->] (X1) -- (X2);

\node[region, fit=(X1)(Y)(X2)] (R) {};

\end{tikzpicture}
}
\caption{}
\label{fig:motivating_example_DAG_b}
\end{subfigure}

\caption{The causal diagram detailing the motivating example. Variables inside the dashed box are used to identify EY-separable functions. Solid edges represent invariant mechanisms while dashed edges represent mechanisms that may vary over environment. We consider two settings with (a) and without (b) hidden confounders.}
\label{fig:motivating_example_DAG}
\end{figure}
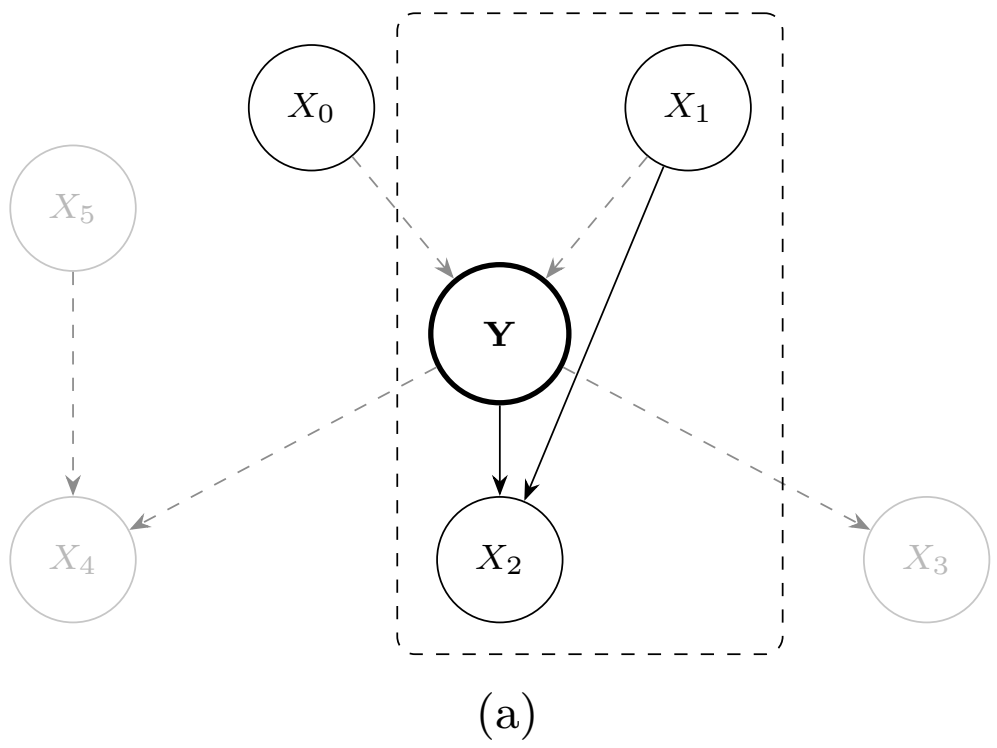

To illustrate EY-separation, we consider a binary classification setting ($L=2$). We consider two toy SCMs starting with the SCM illustrated in Figure~\ref{fig:motivating_example_DAG_a}. The target $Y^e$ has three children and two parents. For the purpose of this example, the nodes $X_0^e$,  $X_1^e$, and $X_2^e$ will be of particular importance. Other nodes may also exist, making the graph increasingly complex.

The target $Y^e$ is generated from a binomial logistic regression model,
\begin{align}
    \P(Y^e = 1\given (X_0^e, X_1^e)=(x_0,x_1)) &= \frac{\exp(\theta_0^e x_0 + \theta_1^e x_1)}{1+ \exp(\theta_0^e x_0 + \theta_1^e x_1) }, \label{equ::toy}\\
    \P(Y^e = 2\given (X_0^e, X_1^e)=(x_0,x_1)) &= 1 - \P(Y^e = 1\given (X_0^e, X_1^e)=(x_0,x_1)).
\end{align}

In this motivating example, $Y$ has two parent nodes $(X_0^e, X_1^e)$ and one \emph{invariant} child node $X_2^e$ (defined below) along with other general child nodes. Other nodes may also exist making the graph increasingly complex, as illustrated in Figure~\ref{fig:motivating_example_DAG}. For the purpose of identifying EY-separable functions, we will see that the variables $X_1^e$ and $X_2^e$ can be used. We shall also see that other variables can be used in addition to $X_1^e$ and $X_2^e$ that lead to EY-separable functions. 

 We assume that for $X_2^e$, the functional form of $f^e_2(\cdot, \cdot)$ varies with $Y^e=y$ such that 
\begin{equation}\label{equ::Xi_def}
X_2^e \leftarrow g_y(X^e_1)  + \epsilon_2^e\; ,
\end{equation}
 where it is crucial that $g_y(\cdot)$ does not depend on $e$, for each $y\in[1:L]$, and that the noise term $\epsilon_2^e$ has zero mean. It is important to note that all the other child nodes are generated following the general assignment $f^e_i (\cdot, \cdot)$ for $i\ne 2$ as in~\eqref{equ::SCM}, which can vary arbitrarily over environment. 

 Now we turn to the prediction task in the test environment $t$, where $Y_t$ is, of course, not available. At first glance, it seems unclear how to decompose the logistic function in~\eqref{equ::toy} into two EY-separable functions as $\theta^t_y$ depends on both $t$ and $y$. Let $S = \{1\}$. The following observation can be made in the test environment where we can apply the law of total probability to get 
 \begin{align*}
        \E[X^t_2\given X_S^t=x] &= \sum_{y=1}^2 \E[X^t_2\given X_S^t=x,Y^t=y]\cdot\P(Y^t=y\given  X_S^t=x) \\
        &= \sum_{y=1}^2 g_y(x)\cdot\P(Y^t=y\given  X_S^t=x). \numberthis\label{equ::examp::dIMP}
    \end{align*}
Writing this in matrix form, we have 

\begin{equation} \label{equ::cond_conf_mat}
\begin{bmatrix}
        \E[X_2^t\given X_{S}^t=x] \\[4pt]
        1
\end{bmatrix} 
    = V
    \times \begin{bmatrix}
        \P(Y^t=1\given  X^t_{S}=x) \\[4pt]
        \P(Y^t=2\given  X^t_{S}=x)
\end{bmatrix},
\end{equation}
where the matrix $V$ does not depend on environment 
\begin{equation} \label{equ::cond_conf_mat_C}
V:= \begin{bmatrix}
        g_{y=1}(x)  & g_{y=2}(x)\\[4pt]
        1  & 1
\end{bmatrix}.
\end{equation}
In generic cases, $g_{y=1}(x)$ and $g_{y=2}(x)$ are different, which means $V$ is invertible; we will make this explicit in Assumption~\ref{assumption1} (also see a variation of this example in Section~\ref{sec:invert} and Remark~\ref{rem:exp_invert}.) Thus 
\begin{equation}  \label{equ::mot::prob_y_is_psiphi}
\P(Y^t=1\given  X^t_{S}=x)
    = V^{-1}\biggl|_{\text{first row}}
\begin{bmatrix}
        \E[X_2^t\given X_{S}^t=x] \\[4pt]
        1
\end{bmatrix}.
\end{equation}
It can then be said that $\rho_y$ and $\psi^t$ are EY-separable since 
\begin{align} \label{equ::motivating_phi_psi}
    \rho_{y=1}(x,w) := V^{-1}\biggl|_{\text{first row}}\cdot \;w \quad\text{ and }\quad
    \psi^t(x) := \begin{bmatrix}
        \E[X_2^t\given X_{S}^t=x] \\[4pt]
        1
\end{bmatrix}. 
\end{align}
Because of this separation, each quantity can be estimated separately: 
\begin{center}
    $\rho_y(x,w)$ can be estimated using only the training environments $e\in \Ec_{\text{train}}$, while $\psi^t(x)$ can be evaluated in the test environment $e=e^{\text{test}}$.
\end{center}
 Then $\P(Y^t=1\given  X^t_{S}=x)$ can be estimated, leading to the classification result. 

\medskip
In the example above, estimating $\P(Y^t=1\given  X^t_{S}=x)$ relies on the subsets of predictors indexed by $S = \{1\}$ and $2$. The choice of these variables is critical in estimating $Y^t$. Selecting other variables in place of $S$ or the child node $X_2^e$ may help or harm in various ways.

\smallskip
\noindent\textbf{Functions are not EY-separable for any $\mathbf{S \subseteq [1:m]}$:}  Suppose that $S =\{5\}$ (see Figure~\ref{fig:motivating_example_DAG}). Then $\E[X^t_2\given X_{S}^t=x,Y^t=y]$ does not describe the parent-child relationship in~\eqref{equ::Xi_def} and will in general vary over environments. The matrix $V$ will then depend on environment and $\rho_y$ and $\psi^e$ will not be EY-separable. Thus, the selection of the predictors $X^e_S$ and $X^e_2$ is critical in estimating $\P(Y^t=1\given  X^t_{S}=x)$. 



\smallskip
\noindent\textbf{EY-separation can be found using various choices for $S$:} While $S = \{5\}$ does not produce EY-separable functions, $S = \{1\}$ is not the only set that can be used to identify EY-separable functions. Suppose that $S = \{1,5\}$. The relationship between $X_5^e$ and other variables is not explicitly defined, but its causal relationship relative to other nodes in the graph is outlined in Figure~\ref{fig:motivating_example_DAG}. We note that the form of~\eqref{equ::examp::dIMP} remains the same as  
\begin{equation} \label{equ::examp::cond_indep}
        \E[X^t_2\given X_S^t=(x_1,x_5),Y^t=y] = \E[X^t_2\given X_1^t=x_1,Y^t=y] = g_y(x_1),
\end{equation}
which follows as $X_5^t \independent X_2^t \given X_1^t,Y^t$. The rest of the example follows similarly, and EY-separable functions can be identified. See Proposition~\ref{prop::suff2} (Condition $2$) for details relating to how inclusive $S$ can be.  

\subsubsection{Variation 1: On the invertibility of $V$}
\label{sec:invert}
To see why we expect $V$ to be invertible in generic cases, we consider a slight variation from our motivating example. Suppose that $L=3$ and $Y$ has an additional \emph{invariant} child node, say $X_3$ with
\begin{equation}\label{equ::X3_def}
X_3^e \leftarrow h_y(X^e_1)  + \epsilon_3^e\; .
\end{equation}
That is, in Figure~\ref{fig:motivating_example_DAG_a}, both $X_1$ and $Y$ will have solid arrows pointing toward $X_3$. Then the same arguments lead to the corresponding $V$
\begin{equation} \label{equ::cond_conf_mat_C}
V:= \begin{bmatrix}
        g_{y=1}(x)  & g_{y=2}(x)  & g_{y=3}(x)\\[4pt]
        h_{y=1}(x)  & h_{y=2}(x)  & h_{y=3}(x)\\[4pt]
        1  & 1 &1
\end{bmatrix}.
\end{equation}
The invertibility thus comes from two sources of variation: the difference between the generating mechanisms of the two child nodes (i.e., $g_y$ vs. $h_y$), and the variation introduced by different class labels, $y\in \{1,2,3\}$. The latter is potentially highly nonlinear due to the nature of classification, even though invertibility holds under linear relationships as well. 

\subsubsection{Variation 2: With hidden confounder}

 We briefly examine the alternative setup in Figure~\ref{fig:motivating_example_DAG_b} to illustrate both necessary and useful properties of EY-separation. In this setting, we assume that $Y^e$ is still generated by a logistic model. However, we assume there exists a hidden confounder $H^e$. Rather, $Y^e$ has two causal parents $H^e$ and $X_1^e$ such that 
\begin{align}
    \P(Y^e = 1\given (X_1^e, H^e)=(x,z)) &= \frac{\exp(\theta_1^e x + \theta_2^e z)}{1+ \exp(\theta_1^e x + \theta_2^e z) }. \label{equ::toyb}
\end{align}

The same process shown above holds even when we introduce the hidden confounder $H^e$. That is, EY-separable functions can be identified that estimate $\P(Y^e = 1\given X_S^e=x)$ for $S=\{1\}$ using the same process that was discussed in the previous example. As neither example leverages $X_0^e$ nor $H^e$, the process remains the same.

\section{Sufficient Conditions under SCMs}
 \label{sec:: EY_SCM}
 
We now examine general conditions that allow $\rho_y$ and $\psi^e$ to be EY-separable under the SCMs~\eqref{equ::SCM}. Recall that $Y^e$ can take on $[1:L]$. As in our motivating example, we will find it necessary that the target $Y^e$ have child nodes, the number of which is dependent on $L$. For each of these children, the generating mechanism must not be intervened, as we will soon show (a practical setting with one child node is discussed in Section~\ref{sec::parm}).

For the pair $(S\subseteq [1:m], C\subseteq [1:m]\backslash S)$, we define the following functions $\rho^{e}_{y,C,S}$ and $\psi_{C,S}^e$ in a similar fashion as the EY-separable functions of~\eqref{equ::motivating_phi_psi} in the motivating example. For simplicity of presentation, we will suppress the subscripts when the underlying $(C, S)$ is clear from the context. First, let $V^e(x):=V^e_{C,S}(x)$ be an $L\times L$ matrix such that the $(c,y)$-th entry of $V^e(x)$ is 
\begin{equation}\label{equ::ve_def}
    v^e(x,y):=v^e_{c,S}(x,y):=\E[X^e_{c}\given X_{S}^e=x,Y^e=y]
\end{equation}
for $c\in [1:L-1]$, while the last row of $V^e(x)$ is a row of $1$'s. The matrix  $V^e(x)$ is a generalization of~\eqref{equ::cond_conf_mat_C} in the motivating example. 


\medskip
As in the motivating example along with the variation in Section~\ref{sec:invert}, we expect the following assumption to hold in generic cases.
\begin{assumption}
\label{assumption1}
Assume that  $V^e(x):=V^e_{C,S}(x)$ is invertible for all $(C,S)$ and $x$. 
\end{assumption}


\begin{remark}
As will be further illustrated in Section~\ref{sec::exp_sim}, the invertibility of $V^e(x)$ can arise from the dependence of the child generating mechanisms on the label $Y^e=y$, even when these mechanisms are linear in the predictors (see Remark~\ref{rem:exp_invert} for details). 
\end{remark}

Based on this, we can then define
\begin{equation}\label{equ::rhoCP_def}
    \rho^{e}_{y}(x,w):=\rho^{e}_{y, C, S}(x,w) := (V^e(x)^{-1})_{y}\cdot w,
\end{equation}
where $(V^e(x)^{-1})_{y}$ denotes the $y$-th row of $V^e(x)^{-1}$. 
Next, we introduce 
\begin{equation}\label{equ::psiCP_def}
    \psi^e(x):=\psi_{C,S}^e(x) := \biggl[\E[X^e_{c_1}\given X_{S}^e=x],\dots,\E[X^e_{c_{L-1}}\given X_{S}^e=x],\,1\biggr]^\top.
\end{equation}
This leads to the following proposition providing sufficient conditions for EY-separation with simple causal interpretations. Let $\text{pa}(X^e_C) := \bigcup_{c\in C} \text{pa}(X^e_c)$. 
\begin{proposition} \label{prop::suff}
     Following the SCMs in~\eqref{equ::SCM} and under Assumption~\ref{assumption1}, if 
    \begin{enumerate}
        \item For each $c\in C$, $f^e_c(\cdot, \cdot)\equiv f_c(\cdot, \cdot)$ with  $\epsilon^e_c\sim \mathsf{F}_c$,  and $Y^e \in \text{pa}(X^e_c)$, \label{itm::invar_f} 
        \item $X_S^e :=  \text{pa}(X^e_C)\backslash\ Y^e $, \label{itm::Xp_comp} \label{itm::nondes_Y}
    \end{enumerate}
    hold true for some pair $(C,S)$, then $\rho^{e}_y\equiv\rho^{y}$ and $\psi^e$ are EY-separable.
\end{proposition}

Instead of proving this directly, we state and prove a slightly more general result, where $X_S^e$ is a set of non-descendants of $Y^e$ that contains $\text{pa}(X^e_C)\backslash\ Y^e$. 
\begin{proposition}
     \label{prop::suff2}
    Following the SCMs in~\eqref{equ::SCM} and under Assumption~\ref{assumption1}, if
    \begin{enumerate}
        \item For each $c\in C$, $f^e_c(\cdot, \cdot)\equiv f_c(\cdot, \cdot)$ with  $\epsilon^e_c\sim \mathsf{F}_c$, and $Y^e \in \text{pa}(X^e_c)$, \label{itm::invar_f} 
        \item $X_S^e := X^e_Q \cup \text{pa}(X^e_C)\backslash\ Y^e  \subseteq \text{nd}(Y^e)$, \label{itm::Xp_comp} \label{itm::nondes_Y}
    \end{enumerate}
     hold true for some pair $(C,S)$, then $\rho^{e}_y\equiv\rho^{y}$ and $\psi^e$ are EY-separable. 
\end{proposition}



In the proof of Proposition~\ref{prop::suff2}, one identity in~\eqref{equ::dimp} plays a critical role, and we will refer to this as the {\bf discrete invariant matching property (dIMP)}. Specifically, a pair $(C,S)$ satisfies dIMP over $e\in \Ec$ if   
        \begin{align} \label{equ::dimp_def}
            \E[X_c^e\given X_S^e=x] &= \sum_{y=1}^L v_{c,S}(x,y)  \cdot \P(Y^e=y\given X_S^e=x),\qquad \text{ for all }c\in C,
        \end{align}
where $v_{c,S}$ does not depend on $e\in\Ec$. 
In other words, dIMP is a sufficient condition for EY-separation. (A preliminary result was reported in~\cite{goddard2024mining}.) Proposition~\ref{prop::suff2} can be written in terms of dIMP as follows.  
\begin{corollary}
\label{coro:dimp}
    If the pair $(C, S)$ satisfies the dIMP over all $e\in\Ec$, then we can identify EY-separable functions as in Proposition~\ref{prop::suff2}. 
\end{corollary}

With Proposition~\ref{prop::suff2} and Corollary~\ref{coro:dimp} in place, we now have explicit constructions demonstrating the power of partial invariance as alluded to before in~\eqref{eq::EY_estimate}. That is, when $(C,S)$ satisfies the conditions in Proposition~\ref{prop::suff2}, then we can estimate $\P (Y^t\given X^t_S)$ because,
\begin{enumerate}
    \setlength\itemsep{0em}
    \item The vector $\psi^t(X_S^t)$, which consists of the expectations $\E[X^t_c\given X^t_S]$, can be estimated using the available predictors in the test environment, 
    \item $V(x)$, which is composed of the expectations $v(x,y)$, does not depend on $e$ and can be estimated using the pooled data over all training environments.
\end{enumerate}

\begin{remark}

This dIMP condition can be viewed as a discrete analogue of the IMP introduced in~\cite{du2023learning} (see in particular in~\cite[Equation~$(11)$]{du2023learning}). Following the IMP, the expected value of $Y^e$ given some subset of variables is expressed as a linear combination of some matching parameters that remain invariant across environments and a combination of some subset of observed predictors at test time. Similarly to the proposed approach here, a portion of the result (the invariant matching parameters) must be estimated at training time, while the final result can only be obtained with an additional regression on only test data. The main distinction is that IMP is formulated as a relation between linear predictors for a continuous response, whereas the present relation is matrix-valued, requires more than one child, and applies to the class probability (not expectation) of a discrete response.
    
\end{remark}

\begin{remark}
    It is possible to identify other sufficient conditions than those presented in Proposition~\ref{prop::suff}. We note the proof would follow similarly if Condition~\ref{itm::Xp_comp} were altered such that $X^e_S = nd(X^e_c)$ for all $c$.
We will see, however, that the condition as stated in Proposition~\ref{prop::suff} will be important when developing an estimator of $Y^t$ (see Section~\ref{sec::methods}).
\end{remark}

We now refer back to the motivating example containing the confounder $H^e$. We note that the sufficient conditions in Propositions~\ref{prop::suff} and~\ref{prop::suff2} do not consider all causal parents of $Y^e$. Thus, if one parent were a hidden confounder, it would have no effect on the estimation of $Y$ (unless it was also a parent of a variable in $C$). Also note that if $H^e$ is a parent node of $Y^e$, it cannot be in $C$ because of the underlying acyclic graph. We note, however, that while Condition~\ref{itm::Xp_comp} of Proposition~\ref{prop::suff2} states that $X^e_S\subseteq nd(Y^e)$, it would never be the case that $X^e_S=nd(Y^e)$ as $H^e\in pa(Y^e)$. 


\begin{remark}   
\label{rem:ey_gen_mech}
Condition~\ref{itm::Xp_comp} in Proposition~\ref{prop::suff2} states that $S$ must index at least the parents of all children not including $Y^e$ and at most any non-descendant of $Y^e$. In the special case of $X_S^e = nd(Y^e)$ (that satisfies the sufficient conditions),  $\E[Y^t\given X_S^t]$ describes the generating process of $Y$ in~\eqref{equ::SCM}, because 
\begin{align*}  \E[Y^t\given X_S^t] &\overset{(a)}{=} \E[Y^t\given X_Q^t,\text{pa}(X^t_C)\backslash Y^t] \\
        &= \E[Y^t\given \text{pa}(Y^t), X_Q^t\backslash \text{pa}(Y^t),\text{pa}(X^t_C)\backslash Y^t] \\
        &\overset{(b)}{=} \E[Y^t\given \text{pa}(Y^t)]=\E[f_Y^t(\text{pa}(Y^t),\epsilon^t_Y)] ,\numberthis\label{equ::PYgXp}
\end{align*}
where $(a)$ follows from $X_S^e = X^e_Q \,\cup\, \text{pa}(X^e_C)\backslash\ Y^e$ (Condition~\ref{itm::nondes_Y}), and $(b)$ follows as $X_Q^t\backslash \text{pa}(Y^t),\text{pa}(X^t_C)\backslash Y^t$ are nondescendants of $Y^t$ (Condition~\ref{itm::nondes_Y}). In our motivating example, this would be the setting where $S = \{0,1\}$ indexes the parents of $Y^e$ and $\P(Y^e\given X_S^e)$ describes the logistic model that generates $Y^e$. 
\end{remark}

\section{Methods and Analysis}\label{sec::methods}

Recall that for a fixed $(C,S)$, $\rho_y$ and $\psi^e$ are EY-separable given the sufficient conditions in Proposition~\ref{prop::suff2}. To begin our discussion, we assume that we have such a $(C,S)$ already identified. We will discuss how to select $(C,S)$ in practice in Section~\ref{sec::ident_CP}, .

Let $n^e$ be the number of samples observed for environment $e\in\Ec$. For the test environment $t\in\Ec$, we are given the unlabeled data $\Xv^t:= [x_1^t,\dots,x_{n^t}^t]$, where each $x_i^t$ is distributed according to $X^t$. For all other environments $e\in\Ec_{\text{train}}$, we are given the labeled data $\Xv^e:= [x_1^e,\dots,x_{n^e}^e]$ and $\Yv^e:= [y_1^e,\dots,y_{n^e}^e]$ distributed according to $X^e$ and $Y^e$, respectively. We denote $\Xv$ as the dataset containing pooled data from $\Xv^e$ for each $e\in\Ec_{\text{train}}$, and do similarly for $\Yv$. For each $i\in n^t$, let $x_{i,S}^t$ be the be the sub-vector of $x_i^t$ indexed by the set $S$. 

 Assuming the conditions in Proposition~\ref{prop::suff2} hold, we propose an estimate of $Y^t$ based on  $\rho_y(x,w)$ and $\psi^e$ defined in~\eqref{equ::rhoCP_def} and~\eqref{equ::psiCP_def}, respectively. To do so, compute an estimate $\widehat{v}$ of $v$ (i.e., the expectation $\E[X^e_c\given Y^e,X^e_S]$) using $\Xv$ and $\Yv$. Note that this estimate is computed on pooled data over all training environments. This can be reasonably done as $v$ does not depend on $e$ under the conditions in Proposition~\ref{prop::suff2} (see~\eqref{equ::p_suf_XcgXpY}). Thus, pooling environments provides a single estimate that can be used for every environment. For $x\in\Xc_S$, let
 \begin{equation}\label{equ::V_hat}
   \widehat{V}(x) := \widehat{V}_{C,S}(x) := (\widehat{v}_{c,S}(x,y))_{c\in C,y\in\Yc} \quad\text{ and }\quad  \widehat{\rho}_y(x,w) := \widehat{\rho}_{y,C,S}(x,w) := (\widehat{V}_{C,S}(x)^{-1})_y \cdot w,
 \end{equation}
 where, like $V(x)$, the last row of $\widehat{V}(x)$ is ones. Then, for each $c\in C$, compute an estimate $\widehat{\psi}^t$ of $\psi^t:=\psi^t_{c,S}$ (i.e., $\E[X_c^t|X_S^t]$) using $\Xv^t$, and let 
 \begin{equation}\label{equ::psi_hat}
     \widehat{\psi}^t(x) := (\widehat{\psi}^t_{c,S}(x))_{c\in C}, 
 \end{equation}
 where the last element in $\widehat{\psi}^t(x)$ is one. We note that as the functions $V$, $\rho$, and $\psi$ all depend on $(C,S)$, so do their estimates $\widehat{V}$, $\widehat{\rho}$, and $\widehat{\psi}$. Again, we often drop $(C,S)$ to simplify notation. We then estimate $Y^t$ using
\begin{equation}\label{equ::meth_multi::class}
    h_{C,S}(x) := \arg\max_{y} \; \widehat{\rho}_y(x,\,\widehat{\psi}^t(x)).
\end{equation}
for any $x\in\Xc_S$. 
A summary of the procedure used to obtain the estimator $\hat{h}_{C,S}$ is as follows: $(a)$ Compute an estimator of $V$ called $\widehat{V}$ and use it to form $\widehat{\rho}$. $(b)$ Compute an estimator of $\psi^t$ called $\widehat{\psi}^t$ and together with $\widehat{V}$, form $\widehat{\rho}_y(x_S,\,\widehat{\psi}^t(x_S))$. $(c)$ Take the largest of this quantity for any $y$ to obtain the estimate $h_{C,S}(x)$. This process is a subroutine of Algorithm~\ref{alg::WdIMP}, which is detailed below. Assuming that the pair $(C,S)$ satisfies the dIMP, $\rho_y$ and $\psi^t$ are EY-separable functions and $\widehat{\rho}_y(x,\widehat{\psi}^t(x))$ will approximate $\P(Y^t \given X^t_S)$. The exact means of estimation of the expectations $\E[X_c^e|X_S^e]$ and $\E[X_c^e|Y^e,X_S^e]$ may vary. In a non-parametric approach, a flexible method may accurately estimate $X^t_c|X^t_S$. In this work, we examine linear and neural network estimators for these expectations as seen in Section~\ref{sec::experiments}.

\subsection{Identifying Suitable $\textbf{(C,S)}$ Pairs}\label{sec::ident_CP}

Using dIMP, we can obtain a classifier, denoted by $h_{C,S}$, that is sufficient to estimate $Y^t$ when the pair $(C,S)$ is known to satisfy the dIMP. As a suitable pair $(C,S)$ is generally unknown, one must be identified methodologically. A common approach in deriving a classifier would be to minimize the risk over all training environments. Its not guaranteed, however, that a classifier trained in this manner will simultaneously minimize the test risk. We propose to rely on the EY-separation risk, defined in~\eqref{equ::EY_risk}. 


Assuming  $\rho_y$ and $\psi^e$ are EY-separable,~\eqref{equ::EY_to_R} shows that minimizing the target EY-separation risk will in turn minimize the risk over the target environment. Similarly, if a pair $(C,S)$ obeys~\eqref{equ::dimp} that in turn guarantees EY-separation, we can also have $\R_{\text{EY},S}^t(h_{C,S}) = \R_S^t(h_{C,S})$. Here we make the dependence of $h$ on $(C,S)$ explicit. In general, for any fixed $(C,S)$ that may not obey dIMP, we can adapt the notation from~\eqref{equ::EY_risk} and write the general form of target risk $\widetilde{\R}_S^t(h_{C,S})$. Therefore, we define the pair $(C^*,S^*)$ that minimizes the target EY-separation risk while satisfying dIMP as, 
\begin{align*}
(C^*,S^*):=&\arg\min_{\substack{C, S}} \widetilde{\R}^t_{\text{EY},S}(h_{C,S}),\quad\text{s.t. $(C,S)$ satisfies dIMP} .
    \numberthis \label{equ::opt_wdimp_finite_hat} 
\end{align*}
The optimization above reflects the population case. In practice, and as we shall see below, one can only assess the dIMP (i.e., the invariance of $\widehat{v}$) over training environments. From a causal standpoint, this is akin to enforcing the conditional invariance of $X_C^e|\text{pa}(X_C^e)$ over all training environments as well as the nondescendant condition of $X^e_S$ in while also minimizing the test risk. 

\begin{remark}
    The dIMP itself does not identify a unique $(C,S)$ pair, as illustrated in the motivating example in Section~\ref{sec::mot_examp}. In general, multiple $(C,S)$ pairs may satisfy the dIMP. Among these valid pairs, the optimization in~\eqref{equ::opt_wdimp_finite_hat} selects the pair that minimizes the EY-separation risk, which, under the dIMP, is equal to the test risk defined in~\eqref{equ::risk}.
\end{remark}

\subsection{Finite-Sample Analysis}
Recall the datasets $\Xv^e$,  $\Yv^e$ and $\Xv^t$ defined in Section~\ref{sec::methods}. For the finite-sample setting, we define the empirical risk of $\widetilde{\R}_S^t(h_{C,S})$ with respect to $n^t$ test samples $\{x_i\}_{i=1}^{n^t}$ for $x_i\in \Xc_S$ as
\begin{align}\label{equ::def_Rhat}
    \widehat{\R}_S^t(h_{C,S}): = 1-\frac{1}{n^t} \sum^{n^t}_{i=1}\widehat{\rho}_{h(x_i)}(x_i,\,\widehat{\psi}^t(x_i)\;; \, C,S).
\end{align}

We first discuss the constraint in~\eqref{equ::opt_wdimp_finite_hat}, as it needs to be quantifiable in the finite-sample setting. Based on dIMP, we make the following observation. If dIMP holds, $v_{c,S}(x,y):= \E[X_c^e|X_S^e=x,Y^e=y]$ does not vary over environment, which translates to $|\widehat{v}^{e_1}_{c,S}(x,y) - \widehat{v}^{e_2}_{c,S}(x,y)| = 0$ for all $(e_1,e_2,x,y)$ in the finite-sample regime. In the causal sense of Proposition~\ref{prop::suff}, this enforces the condition that the generating function $f_c^e$ does not vary over environment. One potential downside with computing $|\widehat{v}^{e_1}_{c,S}(x,y) - \widehat{v}^{e_2}_{c,S}(x,y)| = 0$ for each $(e_1,e_2,x,y)$ is that estimates $h^e_{C,S}$ must be computed for each environment, which depending on the number of environments, could be computationally costly. This potential complexity can be alleviated.  Specifically, if we assume each child variable $X^e_c$ follows an additive noise model, we can take a similar approach to that of the residual distribution test in~\cite{heinze2018invariant}. That is, we estimate the mean of the noise distribution $\epsilon^e_c$ to be $\widehat{\mu}^e_c := \frac{1}{n^e}\sum^{n^e}_{i=1} (x^e_{c,i} - \widehat{v}_{c,S}(x^e_{i,S},y^e_i))$. 
The penalty term $\Lambda_{C,S}$ is introduced to capture this observation. 

The following optimization objective is now proposed as a penalized risk, 
\begin{equation}\label{equ::opt_dimp_finite}
(\widehat{C},\widehat{S}) = \arg\min_{\substack{C,S}} \widehat{\R}_S^t(h_{C,S}) + \Lambda_{C,S}, 
\end{equation}
where the penalty term $\Lambda_{C,S}$ is defined for an arbitrary $(C,S)$ and $\lambda \ge 0$ as 
\begin{equation}\label{equ::pen_Lambda}
     \Lambda_{C,S} := \lambda \sum_{e_1,e_2\in\Ec_{\text{train}}} \sum_{c\in C} |\widehat{\mu}^{e_1}_c - \widehat{\mu}^{e_2}_c|.
\end{equation}
The parameter $\lambda$ controls the tradeoff between the EY-separation risk and the invariance of $\widehat{v}^e$. In practice, $\lambda$ must be selected using some model selection criteria. In this work, we use cross validation over the training environments to select $\lambda$.  We summarize the approach of finding suitable $(C,S)$ in Algorithm~\ref{alg::WdIMP}. From a complexity standpoint, $\Lambda_{C,S}$ is much simpler to compute as the estimate $\widehat{v}_{c,S}$ has already been computed in Step $1$ of dIMP--Classifier. 

\begin{remark}
    Note that this is a relaxation of the condition enforces $|\widehat{v}^{e_1}_{c,S}(x,y) - \widehat{v}^{e_2}_{c,S}(x,y)| = 0$ for all $(e_1,e_2,x,y)$ as we only enforce that the mean of the noise distribution is invariant. Even though $\Lambda_{C,S}$ enforces a weaker condition, it has been shown in practice that a comparison of residual distributions is often sufficient to identify invariances over environment~\cite{peters2011causal,heinze2018invariant,pfister2021stabilizing,du2023learning}.
\end{remark}

\begin{algorithm}
\caption{dIMP--Classifier}\label{alg::WdIMP} 
 \textbf{Input:} $\bm{Y}^e$, for each $e\in\Ec_{\text{train}}$, and $\bm{X}^e$, for each $e\in\Ec$

\textbf{Output:} A selected pair $(\widehat{C},\widehat{S})$ according to~\eqref{equ::opt_dimp_finite}
\begin{algorithmic}
\For{each possible $(C,S)$ pair}
\For{each $c\in C$}
\State Using $\Xv$ and $\Yv$, estimate $\E[X^e_c\given Y^e,X^e_S]$ and call it $\widehat{v}$.
\State Using $\Xv^t$, estimate $\E[X_c^t|X_S^t]$ and call it $\widehat{\psi}^t$.
\EndFor
\State Form the functions $\widehat{V}$, $\widehat{\rho}_y$, and $\widehat{\psi}^t$ as in~\eqref{equ::V_hat} and~\eqref{equ::psi_hat}.
\State  Form the function $h_{C,S}$ using $\widehat{\rho}_y$, and $\widehat{\psi}^t$ according to~\eqref{equ::meth_multi::class}.
\State Compute $s(C,S) = \widehat{\R}_S^t(h_{C,S}) + \Lambda_{C,S}$.
\EndFor
\State Return $h_{\widehat{C},\widehat{S}}(x_{i,S}^t)$ for each $i\in n_t$, where $(\widehat{C},\widehat{S}) = \arg\min_{C,S} s(C,S)$. 
\end{algorithmic}
\end{algorithm}

As the number of possible $(C,S)$ pairs grows exponentially with $m$, Algorithm~\ref{alg::WdIMP} quickly becomes infeasible for a large number of predictors. In such settings, one could search over a random subset of all possible $(C,S)$ pairs. Alternatively, a greedy approach may be adopted to identify a local minima with a small EY-separation risk. Empirically, we have observed that the objective function in~\eqref{equ::opt_dimp_finite} is more sensitive to the choice of $C$ than it is to $S$. Consequently, exhaustively searching over $C$ and greedily searching over $S$ tends to yield a smaller local minima. 

\subsubsection{A Practical Variation: Parametric Modeling}
\label{sec::parm}
Our Assumption~\ref{assumption1} implies that $|C|=L-1$ (recall that the last row of the $V$ matrix consists of all $1$'s).  In practice, when the number of labels is relatively large, this requires we must be able to identify at least $L-1$ invariant children of $Y^e$ in order to form $X_C^e$. In this subsection, we discuss a heuristic approach, allowing us to estimate $U^e$ (as defined in~\eqref{equ::U_S}) using at least one child node. In the case where $|C| \leq L-1$, the system as described in~\eqref{equ::dIMP_mat} becomes underdetermined. In such settings, there is often an infinite number of solutions, only one of which corresponds to the true underlying probability distribution $U^e$. The problem then becomes whether we can get close to recovering the correct solution $U^e$. 
One possible approach is to place a parametric assumption on $U^e$. For example, one can assume that $U^e$ is the popular multinomial logistic model parameterized by $\theta$, i.e., $U^e(x)\equiv U^e(x,\theta)$. The goal is then to find a solution of~\eqref{equ::dIMP_mat} such that $U^e$ is multinomial logistic using 
\begin{equation}\label{equ::single_child_opt}
    \widehat\theta^t_{C,S} := \arg\min_{\theta} \sum_{i=1}^{n^t} \lVert  \widehat{\psi}^t(x_{i}^t)- \widehat{V}(x_{i}^t) \cdot U^t(x_{i}^t;\theta)\rVert^2_2.
\end{equation}
The resulting estimate of $U^t$ is then $U^t(x;\widehat\theta^t)$. The procedure used to estimate $Y^t$ using a single child follows closely to that of the multiple child setting. One still needs to iterate over all $(C,S)$ pairs and find the optimal pairs that minimizes the EY-separation risk. The difference is how we compute the final classifier. In this case, we can define the classifier as $\tilde{h}_{C,S}(x) := \arg\max_{y} \, (U^t(x;\widehat\theta_{C,S}^t))_y$
for any $x\in \Xc_S$. We refer to this  parametric version of dIMP--Classifier as \emph{dIMPparm--Classifier}. 

\subsection{Risk Upper Bounds for dIMP}
We now present the main theoretical results that bound the test risk of our dIMP-based procedure. Consider generic functions $\rho_y$ and $\psi^t$ (not necessarily EY-separable) and their estimates $\widehat{\rho}_y$ and $\widehat{\psi}^t$. We then use $\widehat{\boldsymbol{\rho}} := (\widehat{\rho}_y)_{y\in[1:L]}$ to denote the collection of $\widehat{\rho}_y$ over all $y$. 

We will show that test risk bound can be decomposed into several terms: the EY-separation risk, a term measuring how close functions are to being EY-separable, and terms measuring error in the estimates used. The proof of Proposition~\ref{pro_EY} can be found in Appendix~\ref{apx::T1_proof}
\begin{proposition}\label{pro_EY} 
 Assume that, for all $y$, $\rho_y$ is Lipschitz continuous in the second argument with Lipschitz constant $\eta$. For any fixed $S$, the following target risk bound holds for any classifier $h:\Xc_S\rightarrow \Yc$,
\begin{align*} 
    \R_S^t(h) \leq&  \,\widetilde{\R}_S^t(h;\widehat{\boldsymbol{\rho}},\widehat{\psi}^t) + \underbrace{\max_{\substack{x,y}} | \rho_y(x,\psi^t(x)) - \P(Y^t=y|X^t_S=x)|}_{\text{closeness to EY-separation}}\\
    &\hspace{4em} + \underbrace{\max_{\substack{w, x,y}} |\widehat{\rho}_{y}(x,w) - \rho_{y}(x,w)|  + \eta \max_{x} \lVert\widehat{\psi}^t(x) - \psi^t(x)\rVert_2}_{\text{estimation errors}},\numberthis  \label{equ::t1_risk_bound}
\end{align*} 
where $\lVert \cdot\rVert_2$ denotes the $L^2$ norm, $y\in\Yc$, $w\in\Wc$, and $x\in\Xc_S$.
\end{proposition} 
Proposition~\ref{pro_EY} provides a general bound on the risk of some classifier $h$ in terms of the plug-in estimator of the target risk, the degree in which EY-separation fails, and the estimation errors associated with $\rho_y$ and $\psi^t$. We now consider the performance of the estimate $h_{\widehat{C},\widehat{S}}$ in~\eqref{equ::opt_dimp_finite} compared with the optimal $h_{C^*,S^*}$ in~\eqref{equ::opt_wdimp_finite_hat}. Theorem~\ref{the:2} provides a risk bound that accounts for deviations from invariance across environments, estimation error, and finite sample-complexity. The proof for Theorem~\ref{the:2} can be found in Appendix~\ref{apx::T2_proof}.  For $x\in\Xc_S$, we refer to 
\begin{equation}\label{equ::ey_loss}
    \ell^t(x) := \ell^t(x;\rho_{h(x)},\psi^t) := 1-\rho_{h(x)}(x,\psi^t(x); C,S)
\end{equation}
as the population EY-loss for any $(C,S)$. Let $\Lc_{\mathrm{EY}}^t$ be the of EY-loss function class over all $(C,S)$. 

\begin{theorem} \label{the:2}
    For each $e\in\Ec$, $y\in[1:L]$, and $(C,S)$, assume that $\rho^e_{y,C,S}$ of~\eqref{equ::rhoCP_def} is Lipschitz continuous in the second argument with Lipschitz constant $\eta$. Assume that the EY-loss~\eqref{equ::ey_loss} is uniformly (over $x$) upper-bounded such that $|\ell^t(x)|\leq B$. For any $w\in\mathcal W := \left\{\psi_{C,S}^t(x): x\in\mathcal X_S, S\subseteq [1:m], C\subseteq [1:m]\backslash S \right\}$, also assume $\lVert w\rVert_2 \leq d$. 
    
    Then, under Assumption~\ref{assumption1}, the following target risk bound holds for the dIMP--Classifier and any $e\in\Ec_{\text{train}}$ and with probability at least $1-\delta$ for some $\d\in (0,1)$,
    
    \begin{align*}
        \R_{\widehat{S}}^t&(h_{\widehat{C},\widehat{S}}) \leq \R_{S^*}^t(h_{C^*,S^*}) \\
        & + \underbrace{d\max_{x,y}\, \lVert (V^e_{\widehat{C},\widehat{S}}(x)^{-1})_y-(V^t_{\widehat{C},\widehat{S}}(x)^{-1})_y\rVert_2+ \Lambda_{C^*,S^*} }_{\text{invariance of $V^e$}} + \underbrace{\Gamma_{n^t,\delta}}_{\text{sample complexity}}  \\
        & + \underbrace{2\max_{\substack{x,y\\C,S}}\left( d\lVert (\widehat{V}_{C,S}(x)^{-1})_y-(V_{C,S}^e(x)^{-1})_y\rVert_2 + \eta \lVert \widehat{\psi}_{C,S}^t(x) - \psi^t_{C,S}(x) \rVert_2\right)}_{\text{estimation error}}, \numberthis\label{equ::T2B_res}
    \end{align*}
where $V_{C,S}^e(x)$, $\widehat{V}_{C,S}$, and $\Lambda_{C,S}$ are defined in~\eqref{equ::ve_def},~\eqref{equ::V_hat}, and~\eqref{equ::pen_Lambda} respectively. $\Gamma_{n_t,\delta}$ contains the Rademacher complexity $\mathcal{R}_{n^t}(\mathcal{L}_{\mathrm{EY}})$ over the function class $\mathcal{L}_{\mathrm{EY}}$ such that
\begin{equation}
    \Gamma_{n^t,\delta} := 4\mathcal{R}_{n^t}\!\left(\mathcal{L}_{\mathrm{EY}}\right) + 2B\sqrt{\frac{\log(2/\delta)}{2n^t}}.
\end{equation}
\end{theorem}

\medskip
Theorem~\ref{the:2} bounds the test risk of the dIMP--Classifier estimator into terms associated with invariance, complexity, and estimation error. Two terms involve the invariance over environments of $V^e$. The first, $\lVert (V^e_{\widehat{C},\widehat{S}}(x)^{-1})_y-(V^t_{\widehat{C},\widehat{S}}(x)^{-1})_y\rVert_2$, measures the extend of which the expectations induced by $V^e_{\widehat{C},\widehat{S}}$ vary with respect to the test environment. Assuming the pair $\widehat{C},\widehat{S}$ induces functions that satisfy the sufficient conditions of Proposition~\ref{prop::suff2}, $V^e_{\widehat{C},\widehat{S}}$ will not depend on $e$ and this term will be zero. In general, this difference will be small when the inverse matrices are close across environments. 
The second invariance term, $\Lambda_{C^*,S^*}$, quantifies how effective we are able to measure the dIMP property in the finite-sample setting. Intuitively, this term would be small as the ideal $C^*,S^*$ induces EY-separable functions and $\Lambda_{C^*,S^*}$ is constructed to penalize only functions not EY-separable. The term $\Gamma_{n^t,\delta}$ measures the finite sample complexity of evaluating the EY-separation risk using $n_t$ test samples. In particular, it used the Rademacher complexity as well as an additional term that scales with $1/\sqrt{n_t}$. The remaining two terms quantify the estimation error in estimating $ V_{C,S}^e$ and $\psi^t_{C,S}$ over any $(C,S)$ pair. These two terms are expected to be small when the corresponding conditional expectations are estimated accurately, thanks to the universal approximation of neural networks that we adopt in our experiments.

\section{Experiments}
\label{sec::experiments}
We now test the effectiveness of the proposed dIMP--Classifier approach using real and simulated datasets. For each real and simulated setting, it is assumed that the ground-truth causal model is unknown and must be estimated; i.e., we are unsure of the underlying true $C,S$ pair, and the approach laid out in Algorithm~\ref{alg::WdIMP} must be used. Additionally, the approach in Algorithm~\ref{alg::WdIMP} must be implemented, and methods chosen to model $\E[X^e_c\given Y^e,X^e_S]$ and $\E[X_c^t|X_S^t]$, for given indices $c$ and $S$. In this work, we model these in two ways, using ordinary least squares (OLS), which we denote using LIN, and neural nets, which we denote using NN. The neural nets used are feedforward neural nets with two 20-neuron ReLU hidden layers. In the case of OLS or neural networks, both $\E[X^e_c\given Y^e,X^e_S]$ and $\E[X_c^t|X_S^t]$ are modeled using the chosen function class for any $c$ and $S$.

As a baseline comparison, various classification methods are used, including two causal-based, multi-environment prediction methods. Multinomial logistic regression (LR) and neural nets (NN) are also used. While LR and NN are not expected to perform well, as neither are able to adapt to environmental changes, it serves as a good baseline for determining whether the dIMP--Classifier can adapt to new environments and illustrates the need for more robust multi-environment methods. The implementation of NN follows that of the neural network implementation of the dIMP--Classifier. Two causal-causal based domain adaptation methods are used as comparisons, invariant causal prediction (ICP)~\cite{peters2016causal} and stabilized regression (SR)~\cite{pfister2021stabilizing}. While SR was originally proposed for regression settings, there exist straightforward extensions that allow for use in classification settings, which we implement here using logistic regression in place of OLS. For both ICP and SR these methods, $\alpha = 0.1$. We note that the methods developed in~\cite{chen2022estimating} are based on the reweighting scheme and thus do not provide methods for estimating $\P(Y^t\given X^t_S)$ in the test environment; we thus do not consider these methods in this analysis. 

\subsection{Simulated Experiments}\label{sec::exp_sim}
The data simulation used to illustrate the usefulness of the dIMP--Classifier approach is as follows. We consider the following three settings: 
\begin{enumerate}
    \item $L = 2$ and $|CH|=1$ (as in the motivating example).
    \item $L = |CH|$. For this setting, we employ the dIMP--Classifier Algorithm~\ref{alg::WdIMP}.
    \item $|CH| = 1$. For this setting, we employ the dIMPparm--Classifier Algorithm (see Section~\ref{sec::parm}).
\end{enumerate}
For each setting, we create an SCM consisting of a target $Y$, its parents, and children. The number of parents ($m_{PA}$) is chosen randomly from the interval $[1,5]$. The number of children depends on the setting. Each parent $X_{PA,i}$ is normally distributed such that $X_{PA,i}\sim \Nc(\mu_{PA},1)$, where the mean $\mu_{PA}$ is chosen uniformly from $[1, 2]$ for training environments and $[-2,-1]$ for the test environment. 

Consider the following multinomial logistic model:
\begin{align}
    \P(Y^e = y\given X^e_{PA}=x) &= \frac{\exp(\alpha_y^e x)}{\sum_{y' = 1}^{L}\exp(\alpha_{y'}^e x) }, \quad \text{for $y \in [1:L-1]$}\\
    \P(Y^e = y\given X^e_{PA}=x) &=  1-\sum_{y=1}^{L-1}\P(Y^e = y\given X^e_{PA}=x), \quad \text{for $y=L$}.
\end{align}
The target $Y$ follows a multinomial logistic model where the number of labels $L$ is uniform on $[2,6]$. The logistic coefficients $\alpha^e$ are uniformly random on $[1,2]$ for the training environment and $[-4,0]$ for the test environment. The difference in parameters between test and train is to elicit a joint distribution that is sufficiently different across environments to warrant a domain-adaptation approach. Each child $X^e_{CH_i}$ is linear given $Y=y$. Specifically, 
\begin{equation} \label{equ::sim_lin_CH}
    X^e_{CH_i} = (X_{PA}^e)^\top\beta_{i,y} + \varepsilon_{CH_i}, 
\end{equation}
where $\beta_{i,y}$ is a $m_{PA}\times 1$ vector, and $\varepsilon_{CH_i}\sim \Nc(0,1)$. For each $y\in [1:L]$ and $i \in [1:|CH|]$,  elements in the coefficient vector $\beta_{i,y}$ are sampled randomly from $[-2,2]$. Note that these child coefficients do not change over environment.   
\begin{remark}
\label{rem:exp_invert}
    The dependence of the child generating mechanisms in~\eqref{equ::sim_lin_CH} on the class label $Y=y$ is important for satisfying the invertibility condition in Assumption~\ref{assumption1}. The linearity in $X_{PA}$ alone does provide the needed variation in $V(x)$ for invertibility. Rather, this needed variation is induced by the dependence on the class label $Y=y$. Specifically, since the conditional expectations $\E[X_{CH,i}\mid X_{PA}=x,Y=y_1] = x^\top\beta_{i,y_1}$ and $\E[X_{CH,i}\mid X_{PA}=x,Y=y_2]=x^\top\beta_{i,y_2}$, they will be different in generic cases as  $\beta_{PA_i,y_1}$ and $\beta_{i,y_2}$ are generated randomly for each $y$. Furthermore, $\beta_{i,y}$ is generated randomly for each child node $i$. These two variations leading to invertibility. 
\end{remark}

\begin{figure}
    \centering
    \includegraphics[width=0.5\linewidth]{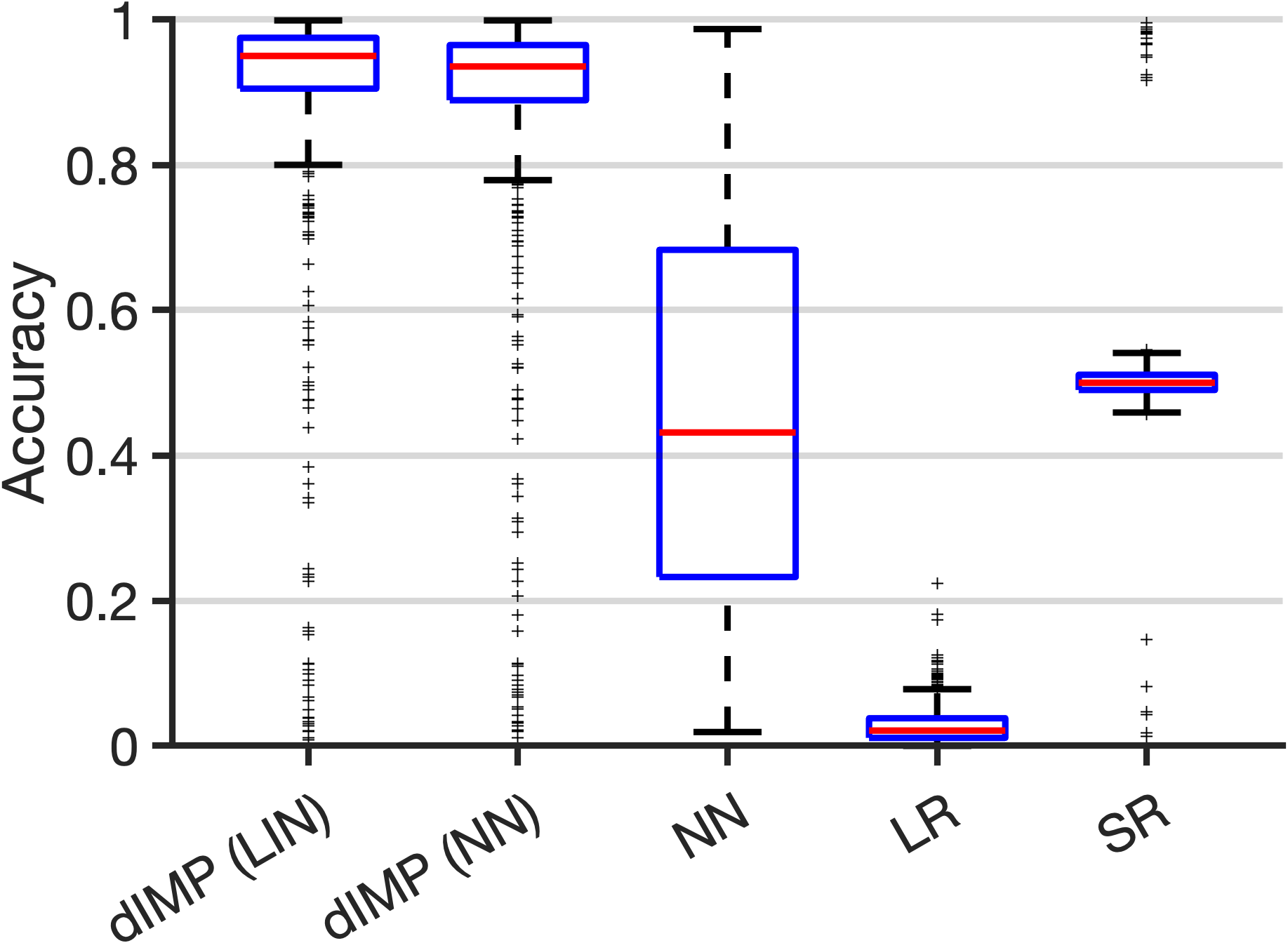}
    \caption{Simulation results for the binary setting when $|CH|=1$.}
    \label{fig::sim_multiCH_bin}
\end{figure}

The first setting considers SCMs where $Y$ is binary and $|CH|=1$. This setting follows similarly to the motivating example in Section~\ref{sec::mot_examp}, where we leverage only a single child and the elements in one of the rows in $V_{C,S}$ are constant. The results for this simulation are presented in Figure~\ref{fig::sim_multiCH_bin}. As can be seen, the dIMP--Classifier implemented using both linear and neural network models achieve average accuracies around $95\%$. The dIMP--Classifier methods outperform all other baselines, with the standard neural network a considerable distance behind them achieving an average accuracy of about $45\%$. As expected, LR performed poorly as it is unable to adapt to the new environment. SR averaged around $50\%$ accuracy in this simulation. We note that in few cases did SR \emph{accept} any subsets, meaning the best it could do was a random guess. 

\begin{figure}[h]
\centering

\begin{subfigure}[]{0.45\linewidth}
  \centering
  \includegraphics[width=\linewidth,height=2.4in,keepaspectratio]{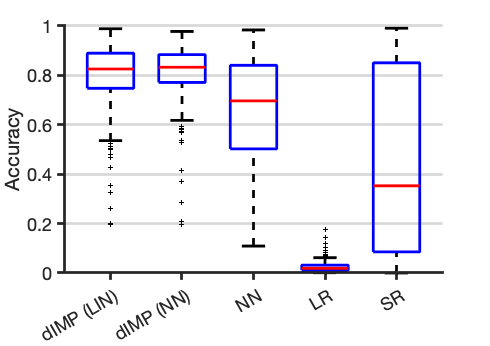}
  \caption{Results for the $|CH|=L$ setting.}
  \label{fig:sim_multiCH}
\end{subfigure}
\hfill
\begin{subfigure}[]{0.45\linewidth}
  \centering
  \includegraphics[width=\linewidth,height=2.4in,keepaspectratio]{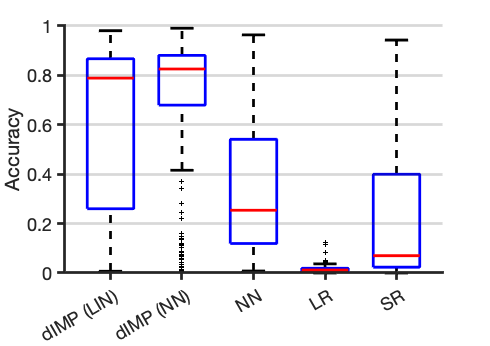}
  \caption{Results for the $|CH| = 1$ setting.}
  \label{sim::sim_singleCH}
\end{subfigure}

\caption{Test accuracy across simulated datasets with a multinomial target.}
\label{fig::light_info}
\end{figure}

The second setting considers SCMs where $|CH| = L$. In these settings, there exist $(C,S)$ pairs with the invertible relationship needed to use the dIMP--Classifier algorithm in~\ref{alg::WdIMP}. We present the accuracy results of dIMP--Classifier compared to the other baseline methods in Figure~\ref{fig:sim_multiCH}. As can be seen, the dIMP--Classifier implemented using both linear models and neural nets outperforms other baseline methods, with the neural network implementation slightly outperforming the linear implementation. Logistic regression is the least performing out of the baseline methods with SR also producing low accuracy. This is to be expected as one of the principal assumptions of SR, that the target node is not intervened, is violated in this simulation. 

The third setting considers SCMs where $|CH| = 1$. In these settings, we use the approach described in Section~\ref{sec::parm}. The accuracy results can be seen in Figure~\ref{sim::sim_singleCH}. Results follow similarly to that of the $|CH| = L$ with some adjustments. While the linear and neural network implementations of the dIMP--Classifier still outperform other baseline methods, their performance does drop when considering only one child. Similarly, the other baseline methods also perform slightly worse than they did in the $|CH| = L$ setting. 


\subsection{Real Experiment: The Light Tunnel Causal Chamber}
The light tunnel causal chamber is a tabletop laboratory that provides datasets from a well-understood optical system~\cite{gamella2025causal} and has been shown to be a reliable, physically grounded benchmark for evaluating prediction under distribution shift and invariance-based methods~\cite{williams2024context,xu2024asymptotically,rahmani2025flow}. Light from an RGB LED is directed through a tube housing two linear polarizers that control the direction of light. Then, three photodiode sensors measure the intensity of the incoming light. The causal graph, as well as the individual variables and their descriptions, can be found in Figure~\ref{fig:light_DAG} and Table~\ref{tbl::light_descriptions}, respectively. Due to the adaptability in controlling parameters such as photodiode size, exposure time, or intensity of RGB values, such a system allows for the flexibility and control to create a wide variety of interventional datasets useful in environment-wise prediction. 

\begin{figure}[h]
\centering

\begin{subfigure}[]{0.425\linewidth}
  \vspace{0pt}
  \centering
  \begin{tikzpicture}[
      baseline=(current bounding box.north),
      >=Stealth,
      every node/.style={font=\footnotesize},
      var/.style={
        circle,
        draw,
        minimum size=1.0cm,
        inner sep=1pt,
        align=center
      },
      target/.style={
        circle,
        draw,
        very thick,
        minimum size=1.1cm,
        inner sep=1pt,
        align=center
      }
    ]

    \node[target] (G) at (0,0) {$\mathbf{G}$};

    \node[var] (V)    at (90:1.8)  {$V$};
    \node[var] (I)     at (210:1.8) {$I$};
    \node[var] (C) at (330:1.8) {$E$};

    \node[var] (t1) at ($(I) + (-1.41, -0.71)$) {$\theta_{1}$};
    \node[var] (t2) at ($(I)+(-1.5,0.5)$) {$\theta_{2}$};
    \node[var] (l1)  at ($(V) + (-1.08, -1.15)$) {$L_{1}$};
    \node[var] (l2)  at ($(V) + (-2.25, -0.5)$) {$L_{2}$};

    \draw[->] (G) -- (V);
    \draw[->] (G) -- (I);
    \draw[->] (G) -- (C);

    \draw[->] (t1) -- (I);
    \draw[->] (t2) -- (I);
    \draw[->] (l1)  -- (I);
    \draw[->] (l2)  -- (I);
    \draw[->] (l1)  -- (V);
    \draw[->] (l2)  -- (V);

  \end{tikzpicture}
  \caption{Ground truth causal DAG.}
  \label{fig:light_DAG}
\end{subfigure}
\hfill
\begin{subfigure}[]{0.5\linewidth}
  \vspace{0pt}
  \centering
  \renewcommand{\arraystretch}{1.2}
  \footnotesize
  \begin{tabular}{ |c||>{\raggedright\arraybackslash}p{0.6\linewidth}| }
    \hline 
     Variable & Description \\ 
     \hline\hline 
     $G$ & Brightness of the green component of main RGB LED \\ 
     \hline
     $V$ & Visible light measurement for the $i$-th intensity sensor \\
     \hline
     $I$ & Infrared measurement for the $i$-th intensity sensor \\
     \hline
     $L_{i}$ & Brightness of the $i$-th LED placed by the light sensor \\
     \hline
     $\theta_i$ & Angle of the $i$-th polarizer frame \\
     \hline
     $E$ & Electric current drawn by the main LED \\
     \hline
  \end{tabular}
  \renewcommand{\arraystretch}{1.0}
  \caption{Variable names and descriptions.}
  \label{tbl::light_descriptions}
\end{subfigure}

\caption{Light tunnel Causal Chamber setup used in the real-data experiments.}
\label{fig::light_info}
\end{figure}

We note that the DAG presented in Figure~\ref{fig:light_DAG} and the subsequent variables used in this analysis are a subset of the $39$ available variables presented in~\cite{gamella2025causal}. We present the variables outlined in Table~\ref{tbl::light_descriptions} for the following reasons. (a) For the chosen environmental experiments, many of the variables were constant. (b) There are several sets of variables in those of~\cite{gamella2025causal} that measured the same physical process but in a slightly different way. For example, two photodiode sensors that are placed only slightly apart. Our analysis includes the largest set of variables where each variable represents a unique physical process while retaining each variable's parent nodes. 

For these experiments, we consider the prediction of the target variable $G$, discretized to three levels. In a purely observational setting, the value of the variable $G$ is chosen uniformly. Interventional datasets have been observed by limiting this uniform distribution such that the value $G$ falls in a \emph{low}, \emph{medium}, or \emph{high} range. We choose these interventions specifically as they intervene on the child node, which is of particular interest in this work, while leaving the children nodes unintervened. 

\begin{figure}[h]
    \centering
    \begin{subfigure}[t]{0.31\textwidth}
        \centering
        \includegraphics[height=4.2cm]{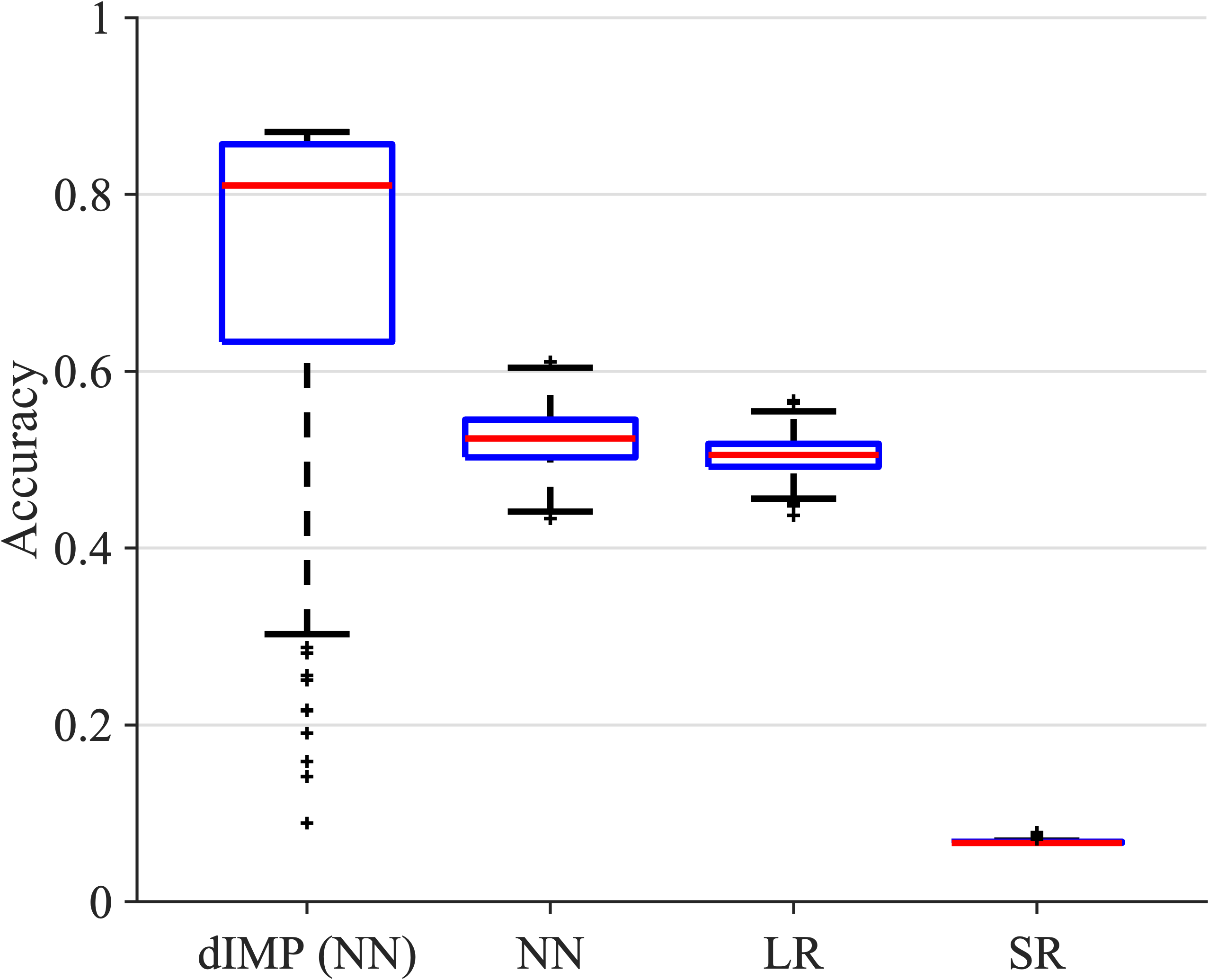}
        \caption{Test Env. - Green Low}
        \label{fig:boxplot_mid}
    \end{subfigure}
    \hfill
    \begin{subfigure}[t]{0.31\textwidth}
        \centering
        \includegraphics[height=4.2cm]{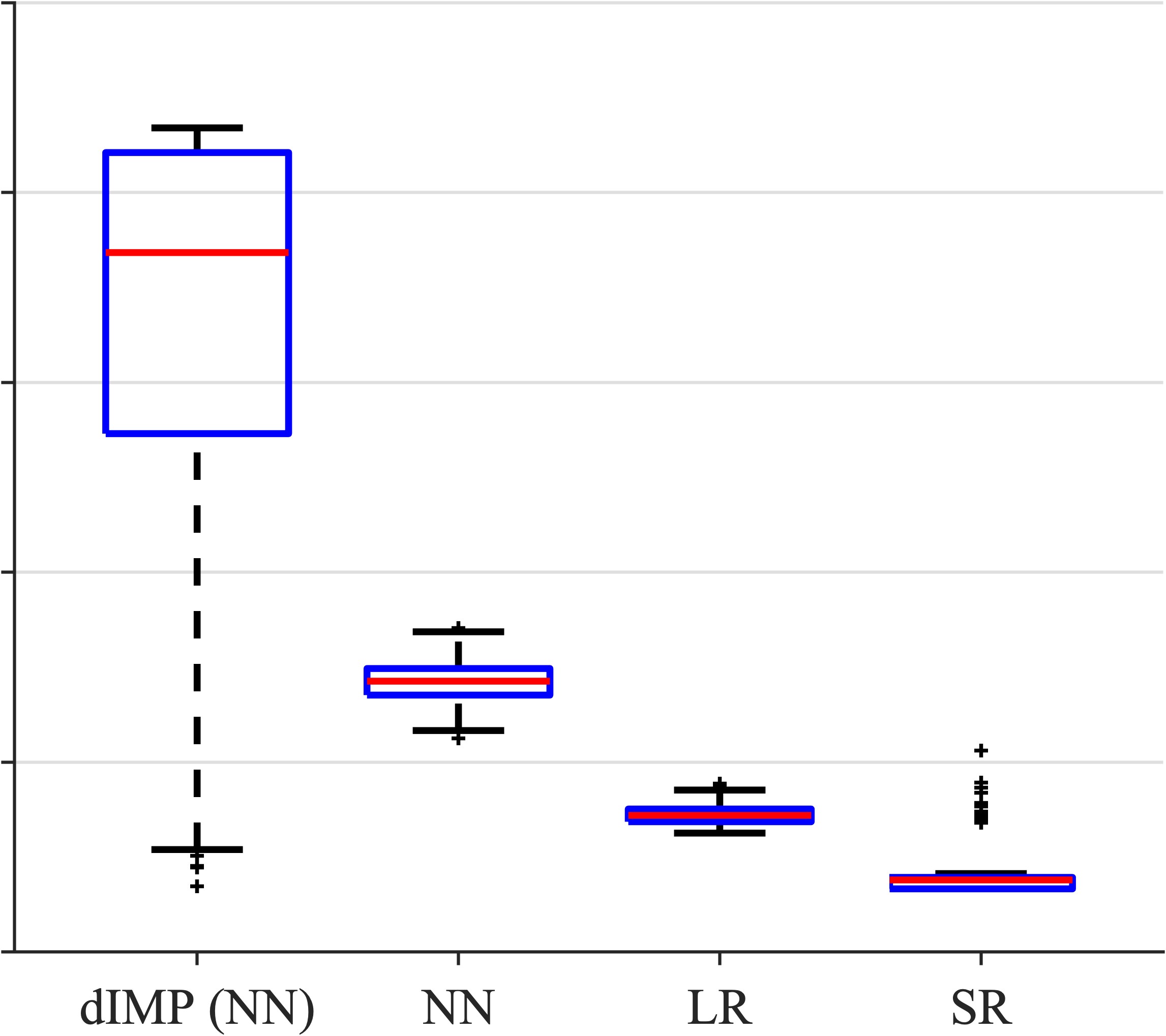}
        \caption{Test Env. - Green Medium}
        \label{fig:boxplot_small}
    \end{subfigure}
    \hfill
    \begin{subfigure}[t]{0.31\textwidth}
        \centering
        \includegraphics[height=4.2cm]{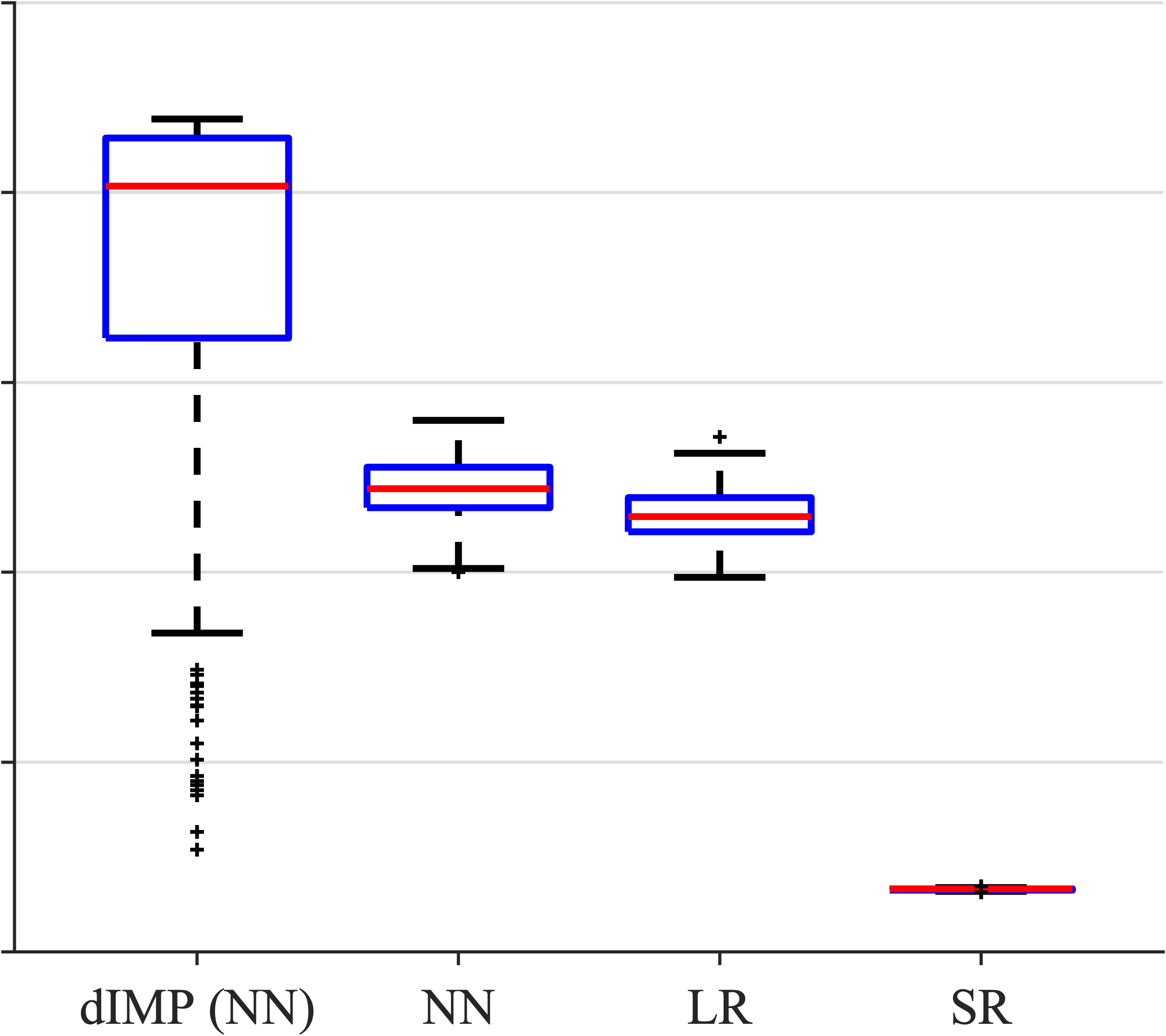}
        \caption{Test Env. - Green High}
        \label{fig:boxplot_large}
    \end{subfigure}
    \caption{Test accuracy results from the light tunnel causal chambers dataset.}
    \label{fig:boxplots_three_envs}
\end{figure}

In the light tunnel causal chamber setting, we consider three environments: two training environments and one test environment. There are three potential options for the test environment. As none of them can be favored over the others, we run the analysis three times, once for each potential test environment. Given two train and one test environment (each $1000$ samples), we take subsamples of $750$ to use as inputs to dIMP--Classifier and other comparison methods. This is repeated $1000$ times to produce the accuracy comparisons in Figure~\ref{fig:boxplots_three_envs}. Over all subsampled experiments, dIMP--Classifier (NN) outperforms all other baseline methods for all environment selections. As expected, SR performs poorly in all environments. NN and LR perform slightly better, achieving an average accuracy of up to  $50\%$ for the \emph{low} and \emph{high} environments and $25\%$ for the \emph{medium} environment. 


\section{Conclusion and Discussion}
\label{sec::discuss}
In this work, we consider classification under distribution shift when the target-generating mechanism varies across environments. We introduce environment-label separation (EY-separation), which relaxes the requirement that the full conditional distribution $\P(Y^e \given X_S^e)$ remain invariant and instead separates it into an environment-invariant component and an environment-dependent component. Within the SCM framework, we provide sufficient conditions for EY-separation, which motivate our main algorithm, dIMP Classifier. We provide finite-sample guarantees showing that test performance depends on deviations from invariance, estimation error, and sample complexity. The empirical results support these conclusions. Across the simulated settings, the proposed classifiers consistently outperform the baseline logistic and neural network methods, as well as invariance-based methods, when the target mechanism is intervened. The real-data experiments on the light tunnel Causal Chamber provide further evidence that the proposed framework is effective in physically grounded settings. Moreover, highly ranked $(C,S)$ pairs frequently reflect the child and spousal relationships predicted by the SCM-based analysis. 

One important future direction is to relax the assumption of exact invariance. In realistic systems, otherwise stable mechanisms may vary slightly across environments due to measurement error, unmodeled effects, or gradual changes in system parameters. Extending EY-separation and dIMP to approximate invariance could therefore lead to guarantees that degrade smoothly with the magnitude of the invariance violation rather than requiring exact invariance.


\newpage
\bibliographystyle{IEEEtran}
\bibliography{ref}

\newpage
\appendix
\section{Proofs} \label{sec::app}
\subsection{Proof of Proposition~\ref{prop::suff2}}
    \begin{proof}
In the proof, we make the dependency on $(C,S)$ explicit. We first show that when the conditions on $(C,S)$ hold, then $\rho_{y, C,S}^{e}$ does not depend on $e$. For an acyclic SCM in~\eqref{equ::SCM}, we make the following two observations: (1) any variable is conditionally independent of all nondescendants given its parents (parental Markov condition~\cite{pearl2009causality}), and (2) the nondescendants of $Y^e$ are also nondescendants its children. Thus, as $X^e_c$ is a child of $Y^e$ (Condition~\ref{itm::invar_f}), $\text{nd}(Y^e) \subseteq \text{nd}(X_c^e)$. Following this, we have $X^e_Q\subseteq \text{nd}(X^e_c)$ because $X^e_Q\subseteq \text{nd}(Y^e)$ (Condition~\ref{itm::Xp_comp}). On the other hand, observe that $\text{pa}(X_C^e)\backslash \text{pa}(X_c^e) \subseteq \text{pa}(X_C^e)\backslash \text{pa}(Y^e)$ and the latter is contained in $\text{nd}(Y^e_c)$ (Condition~\ref{itm::nondes_Y}), which implies 
\[
\text{pa}(X_C^e)\backslash \text{pa}(X_c^e)\subseteq \text{nd}(Y^e_c) \subseteq \text{nd}(X^e_c).
\]
          
          Thus we have that $\text{pa}(X_C^e)\backslash \text{pa}(X_c^e)$ and $X_Q^e$ are all nondescendants of $X^e_c$, and we can write the following due to the parental Markov condition,
         \begin{equation}\label{equ::p_suf_indep}
             \{\text{pa}(X^e_C)\backslash \text{pa}(X_c^e), X_Q^e \}\independent X^e_c\given \text{pa}(X^e_c).
         \end{equation}
        Then, for each $c\in C$
        \begin{align*}\label{equ::p_suf_XcgXpY}
            \E[X_c^e\given Y^e,X_S^e] &\overset{(a)}{=} \E[X_c^e\given \text{pa}(X_C^e),X_Q^e] \\
            &=  \E[X_c^e\given \text{pa}(X_C^e)\backslash \text{pa}(X^e_c), X_Q^e,\text{pa}(X^e_c)]\overset{(b)}{=} \E[X_c^e\given \text{pa}(X^e_c)], \numberthis
        \end{align*}
        where $(a)$ follows from replacing $X_S^e$ with $X^e_Q \cup \text{pa}(X_C^e)\backslash\ Y^e$ (Condition~\ref{itm::Xp_comp}), and by noting that $Y^e\in \text{pa}(X^e_c)$ (Condition~\ref{itm::invar_f}), and $(b)$ follows from~\eqref{equ::p_suf_indep}.
Recall that $\epsilon_c^e\sim \mathsf{F}_c$. Let $x\in \Xc_{\text{pa}(Y^e)}$, $x'\in \Xc_{\text{pa}(X_c^e)\backslash Y^e}$. Now, following~\eqref{equ::p_suf_XcgXpY} 
        \begin{align*}
        v^e_{c,S}((x,x'),y) &= \E[X_c^e\given Y^e=y,X_S^e=(x,x')] 
        \\&= \E[X_c^e\given \text{pa}(X^e_c)=(x',y)]\\
        &\overset{(a)}{=} \E[f_c((x',y),\epsilon_c^e)] = \int f_c((x',y),z) \, d\mathsf{F}_c, \numberthis 
        \end{align*} 
        where $(a)$ follows from $X_c^e = f_c^e(\text{pa}(X_c^e),\epsilon^e_c)$ as in~\eqref{equ::SCM} and $f_c^e(\cdot, \cdot)\equiv f_c(\cdot, \cdot)$ in Condition~\ref{itm::invar_f}. As $f_c$  and $\mathsf{F}_c$ do not vary over environment, neither does $v^e_{c,S}((x,x'),y) \equiv v_{c,S}((x,x'),y)$, nor $V^e_{C,S}((x,x'),y) \equiv V_{C,S}((x,x'),y)$.
        Thus, $\rho^{e}_{y, C,S}(x,w)\equiv\rho_{y, C,S}(x,w)$.

 With the invariant $v_{c,S}$ in place, we can write down a key property that decomposes $\E[X_c^e\given X_S^e=x]$ using the law of total probability,   
        \begin{align} \label{equ::dimp}
            \E[X_c^e\given X_S^e=x] &= \sum_{y=1}^L v_{c,S}(x,y)  \cdot \P(Y^e=y\given X_S^e=x),\qquad \text{ for all }c\in C.
        \end{align}
 We now make use of~\eqref{equ::dimp} to show that $\rho_{y, C,S}(x,\psi^e_{C,S}(x)) = \P(Y^e=y\given X_S=x)$.
        This can be written in a matrix form as
        \begin{equation} \label{equ::dIMP_mat}
            \psi_{C,S}^e(x) =  V_{C,S}(x) \cdot U_S^e(x),
        \end{equation}
        where 
        \begin{equation}\label{equ::U_S}
            U^e(x) := U_S^e(x) := \left[\P(Y^e=1\given X_{S}^e=x),\dots, \P(Y^e=L\given X_{S}^e=x)\right]^\top.
        \end{equation}
        As $ V_{C,S}(x)$ is invertible, we can write $U_S^e(x) = (V_{C,S}(x))^{-1} \cdot\psi_{C,S}^e(x)$ and $ \P(Y^e=y\given X_S=x) = \rho_{y,C,S}(x,\psi^e_{C,S}(x))$. Thus, $\rho_{y,C,S}(x,w)$ and $\psi^e_{C,S}$ are EY-separable as the former does not depend on $e$ (by~\eqref{equ::p_suf_XcgXpY}) and the later does not depend on $y$ (by definition).  
    \end{proof}
\subsection{Proof of Proposition~\ref{pro_EY}}\label{apx::T1_proof}
\begin{proof} 
The target risk can then be decomposed as
    \begin{align*}
        \R_S^t(h) &
        \leq \widetilde{\R}_S^t(h;\widehat{\boldsymbol{\rho}}^t,\widehat{\psi}^t) + |\R_S^t(h) - \widetilde{\R}_S^t(h;\boldsymbol{\rho}^t,\psi^t)|\\
        &\quad+ |\widetilde{\R}_S^t(h;\boldsymbol{\rho}^t,\psi^t) - \widetilde{\R}_S^t(h;\boldsymbol{\rho}^t,\widehat{\psi}^t)| + |\widetilde{\R}_S^t(h;\boldsymbol{\rho}^t,\widehat{\psi}^t) - \widetilde{\R}_S^t(h;\widehat{\boldsymbol{\rho}}^t,\widehat{\psi}^t)| .
    \end{align*}
    We now consider terms individually, starting with the term $|\R_S^t(h) - \widetilde{\R}_S^t(h;\rho,\psi^t)|$,
    \begin{align}
         |\R^t(h) - \widetilde{\R}_S^t(h;\boldsymbol{\rho}^t,\psi^t)| &= \left| \P(Y^t\ne h(X^t)) - \E[1-\rho^t_{f(X_S^t)}(X^t_S,\psi^t(X^t_S))] \right| \\
         &= \left|  \E[\rho^t_{h(X^t)}(X^t_S,\psi^t(X^t_S))] -\P(Y^t = h(X^t))\right| \\
         &= \left| \int_{-\infty}^{\infty} (\rho^t_{f(x)}(x,\psi^t(x)) -\P(Y^t = f(x)|X^t_S=x))\P(X^t_S=x) \, dx \right| \\
         &= \left| \E[\rho^t_{h(X^t)}(X^t_S,\psi^t(X^t_S)) -\P(Y^t = h(X^t)|X^t_S)]\right| \\
         &\leq \max_{x\in \Xc_S,y\in\Yc} \left|  \rho^t_{y}(x,\psi^t(x)) -\P(Y^t = y|X^t_S=x)\right|.  \numberthis\label{equ::t1::dist}
    \end{align}
    For the term $|\widetilde{\R}_S^t(h;\boldsymbol{\rho}^t,\psi^t) - \widetilde{\R}_S^t(h;\boldsymbol{\rho}^t,\widehat{\psi}^t)|$, observe 
    \begin{align*}
        |\widetilde{\R}_S^t(h;\boldsymbol{\rho}^t,\psi^t) - \widetilde{\R}_S^t(h;\boldsymbol{\rho}^t,\widehat{\psi}^t)| &= \left|\E[\rho^t_{h(X^t)}(X^t_S,\widehat{\psi}^t(X^t_S)) -\rho^t_{h(X^t)}(X^t_S,\psi^t(X^t_S))]\right|\\
        &\leq \max_{x\in\Xc_S} \left|\rho^t_{f(x)}(x,\widehat{\psi}^t(x)) -\rho^t_{f(x)}(x,\psi^t(x))\right| \\
        &\overset{(a)}{\leq} \max_{x\in\Xc_S}  \eta\cdot \lVert \widehat{\psi}^t(x) - \psi^t(x) \rVert_2, \numberthis\label{equ::t1::bnd_g}
    \end{align*}
    where $(a)$ holds due to the Lipschitz assumption on $\rho^t_{y}$. Similarly, the term $|\widetilde{\R}_S^t(h;\boldsymbol{\rho}^t,\widehat{\psi}^t) - \widetilde{\R}_S^t(h;\widehat{\boldsymbol{\rho}}^t,\widehat{\psi}^t)|$ can be bounded such that
    \begin{align*}
        |\widetilde{\R}_S^t(h;\boldsymbol{\rho}^t,\widehat{\psi}^t) - \widetilde{\R}_S^t(h;\widehat{\boldsymbol{\rho}}^t,\widehat{\psi}^t)| &\leq \max_{x\in\Xc_S} \left|\widehat{\rho}^t_{f(x)}(x,\widehat{\psi}^t(x)) -\rho^t_{f(x)}(x,\widehat{\psi}^t(x))\right| \\
        &\leq \max_{x\in\Xc_S,y\in\Yc} \left|\widehat{\rho}^t_{y}(x,\widehat{\psi}^t(x)) -\rho^t_{y}(x,\widehat{\psi}^t(x))\right|  \\
        &\leq \max_{\substack{w\in\Wc, x\in\Xc_S\\y\in\Yc}} \left|\widehat{\rho}^t_{y}(x,w) -\rho^t_{y}(x,w)\right|.\numberthis\label{equ::t1::bnd_rho}
    \end{align*}
    The bound in~\eqref{equ::t1_risk_bound} follows by considering all bounded terms together.
\end{proof}

\subsection{Proof of Theorem~\ref{the:2}}\label{apx::T2_proof}
\begin{proof}
For simplicity of presentation denote, $h_{\widehat{C},\widehat{S}}$, $h_{C^*,S^*}$, $\rho^e_{\widehat{C},\widehat{S}}$, $\rho^e_{C^*,S^*}$, $V^e_{\widehat{C},\widehat{S}}(x)$, $\psi^t_{\widehat{C},\widehat{S}}$, and $\psi^t_{C^*,S^*}$ as $h$, $h^*$, $\rho^e$, $\rho^{*}_e$, $V^e$, $\psi^t$, and $\psi^{t,*}$, respectively. Do similarly for their estimates. As done in the proof of Proposition~\ref{pro_EY}, we then denote the EY-separation risk in terms of $\rho_y$ and $\psi^e$ using $\widetilde{\R}_S^t(h;\rho,\psi^t)$ and do similarly for $\widehat{\R}_{\widehat{S}}^t$. Now expand $\R_{\widehat{S}}^t(h)$ such that
\begin{align}
   \R_{\widehat{S}}^t(h)  &= \R_{\widehat{S}}^t(h^*) + \underbrace{
      \R_{\widehat{S}}^t(h)-  \widetilde{\R}_{\widehat{S}}^t(h;\boldsymbol{\rho}^e,\psi^t)
   }_{A} \\
   &\quad +  \underbrace{
       \widetilde{\R}_{\widehat{S}}^t(h;\boldsymbol{\rho}^e,\psi^t) - \widehat{\R}_{\widehat{S}}^t(h;\boldsymbol{\rho}^e,\psi^t)
   }_{B_1} \\
   &\quad +  \underbrace{
       \widehat{\R}_{\widehat{S}}^t(h;\boldsymbol{\rho}^e,\psi^t) - \widehat{\R}_{\widehat{S}}^t(h;\widehat{\boldsymbol{\rho}},\widehat{\psi}^t)
   }_{C_1} \\
   &\quad +  \underbrace{
      \widehat{\R}_{\widehat{S}}^t(h;\widehat{\boldsymbol{\rho}},\widehat{\psi}^t) - \widehat{\R}_{S^*}^t(h^*;\widehat{\boldsymbol{\rho}}^*,\widehat{\psi}^{t,*})
   }_{D} \\
   &\quad +  \underbrace{
     \widehat{\R}_{S^*}^t(h^*;\widehat{\boldsymbol{\rho}}^*,\widehat{\psi}^{t,*}) - \widehat{\R}_{S^*}^t(h^*;\boldsymbol{\rho}_e^*,\psi^{t,*})
   }_{C_2} \\
   &\quad +  \underbrace{
      \widehat{\R}_{S^*}^t(h^*;\boldsymbol{\rho}_e^*,\psi^{t,*}) - \widetilde{\R}_{S^*}^t(h^*;\boldsymbol{\rho}_e^*,\psi^{t,*})
   }_{B_2} \\
   &\quad +  \underbrace{
      \widetilde{\R}_{S^*}^t(h^*;\boldsymbol{\rho}_e^*,\psi^{t,*}) - \R_{S^*}^t(h^*)
   }_{E}.
\end{align}
We now bound each term individually. 

\noindent\textbf{Term $\mathbf{A}$}: Following what was done in~\eqref{equ::t1::dist}, 
    \begin{align*}
        \R_{\widehat{S}}^t(h)-  \widetilde{\R}_{\widehat{S}}^t(h;\boldsymbol{\rho}^e,\psi^t) &\leq \max_{x\in \Xc_S,y\in\Yc} \left| \rho^e_{y}(x,\psi^t(x)) -\P(Y^t = y|X^t_S=x) \right| \\
        &\leq \max_{x\in \Xc_S,y\in\Yc} \left| (V^e(x)^{-1})_y\cdot \psi^t(x) - (V^t(x)^{-1})_y\cdot \psi^t(x) \right| \\
        &\leq \max_{\substack{x\in\Xc_{S},y\in\Yc\\w\in\Wc}} \left| (V^e(x)^{-1})_y\cdot w - (V^t(x)^{-1})_y\cdot w \right|. 
        \numberthis\label{equ::t2::A}
    \end{align*}
Then, as $\lVert w\rVert_2 \leq d$ for any $w\in\mathcal W := \left\{\psi_{C,S}^t(x): x\in\mathcal X_S , P\subseteq [1:m], C\subseteq [1:m]\backslash P\right\}$, and using the Cauchy-Schwarz inequality,~\eqref{equ::t2::A} can be further bounded by 
    \begin{align*}
        \max_{\substack{x\in\Xc_{S},y\in\Yc\\w\in\Wc}} \left| (V^e(x)^{-1})_y\cdot w - (V^t(x)^{-1})_y\cdot w \right| &\leq \max_{\substack{x\in\Xc_{S},y\in\Yc\\w\in\Wc}} \lVert (V^e(x)^{-1})_y - (V^t(x)^{-1})_y \rVert_2 \cdot\lVert w\rVert_2 \\
        &\leq d\max_{\substack{x\in\Xc_{S},y\in\Yc}} \lVert (V^e(x)^{-1})_y - (V^t(x)^{-1})_y \rVert_2
        \numberthis\label{equ::t2::A2}
    \end{align*}
\noindent\textbf{Terms $\mathbf{B1}$ and $\mathbf{B2}$}: 
The term $\mathbf{B1} + \mathbf{B2}$ can be bounded such that
\begin{align*}
&\widetilde{\R}_{\widehat{S}}^t(h;\boldsymbol{\rho}^e,\psi^t) - \widehat{\R}_{\widehat{S}}^t(h;\boldsymbol{\rho}^e,\psi^t) \\
&\quad+\widehat{\R}_{S^*}^t(h^*;\boldsymbol{\rho}_e^*,\psi^{t,*}) - \R_{\text{EY},S^*}(h^*;\boldsymbol{\rho}_e^*,\psi^{t,*}) \\
&\leq |\widetilde{\R}_{\widehat{S}}^t(h;\boldsymbol{\rho}^e,\psi^t) - \widehat{\R}_{\widehat{S}}^t(h;\boldsymbol{\rho}^e,\psi^t)| \\
&\quad+ |\widehat{\R}_{S^*}^t(h^*;\boldsymbol{\rho}_e^*,\psi^{t,*}) - \R_{\text{EY},S^*}(h^*;\boldsymbol{\rho}_e^*,\psi^{t,*})|\\
& \leq \left|\E[\ell^t(X^t_{\widehat{S}};\rho^e_{h(X^t_{\widehat{S}})},\psi^t)]    - \frac{1}{n^t} \sum^{n^t}_{i=1} \ell^t(x^t_{i,\widehat{S}};\rho^e_{h(x^t_{i,\widehat{S}})},\psi^t)\right|\\
&\quad + \left|\frac{1}{n^t} \sum^{n^t}_{i=1}\ell^t(x^t_{i,S^*};\rho^*_{h^*(x^{t}_{i})},\psi^{t,*}) - \E[\ell^t(X^t_{S^*};\rho^*_{h^*(X^t_{S^*})},\psi^{t,*})]\right| \\
& \leq 
2\sup_{\ell^t\in\Lc_{\mathrm{EY}}^t}\left|\mathbb{E}\!\left[\ell^t(X_S^t;\rho_y^e,\psi^t)\right] - \frac{1}{n_t}\sum_{i=1}^{n_t}\ell^t(x_{i,S}^t;\rho_y^e,\psi^t) \right|,\numberthis\label{equ::t_B1B2_1}
\end{align*}
where taking the supremum over $\ell^t$ is equivalent of taking the supremum over $(\rho_y^e,\psi^t)$. Following  of~\cite[Theorem~2.1]{bartlett2021deep}, and since the EY-separation risk is bounded by $B$, the term can be further bounded using the Rademacher complexity over the function class $\Lc_{\mathrm{EY}}^t$, denoted as $\mathcal{R}_{n^t}(\Lc_{\mathrm{EY}}^t)$, by
\begin{equation}\label{equ::t_B1B2_2}
2\sup_{\ell^t\in\Lc_{\mathrm{EY}}^t} \left|
\mathbb{E}\!\left[\ell^t(X_S^t;\rho_y^e,\psi^t)\right] - \frac{1}{n_t}\sum_{i=1}^{n_t}\ell^t(x_{i,S}^t;\rho_y^e,\psi^t) \right|
\leq 4\R_{n_t}\!\left(\Lc_{\mathrm{EY}}^t\right) + 2B\sqrt{\frac{\log(2/\delta)}{2n_t}}
\end{equation}
with probability at least $1-\delta$.

\noindent\textbf{Terms $\mathbf{C_1}$ and $\mathbf{C_2}$}: We start by showing that 
\begin{align*}
&|\widehat{\R}_{\widehat{S}}^t(h;\boldsymbol{\rho}^e,\psi^t) - \widehat{\R}_{\widehat{S}}^t(h;\widehat{\boldsymbol{\rho}},\widehat{\psi}^t)|\\
& = \left|\frac{1}{n^t} \sum^{n^t}_{i=1} \widehat{\rho}_{h(x^t_i)}(x^t_{i},\widehat{\psi}^t(x^t_{i}))-\frac{1}{n^t} \sum^{n^t}_{i=1} \rho^e_{h(x^t_i)}(x^t_i,\psi^t(x^t_i))\right|\\
& \leq \max_{x\in\Xc_S}\left| \widehat{\rho}_{h(x)}(x,\widehat{\psi}^t(x))- \rho^e_{h(x)}(x,\psi^t(x))\right|\\
& \leq \max_{x\in\Xc_S}\left[\left| \widehat{\rho}_{h(x)}(x,\widehat{\psi}^t(x))- \rho^e_{h(x)}(x,\widehat{\psi}^t(x))\right| + \left|\rho^e_{h(x)}(x,\widehat{\psi}^t(x))- \rho^e_{h(x)}(x,\psi^t(x))\right|\right]\\
& \overset{(a)}{\leq} \max_{\substack{x\in\Xc_{S},y\in\Yc\\w\in\Wc}} \left| \widehat{\rho}_y(x,w)- \rho^e_y(x,w)\right| + \eta\max_{x\in\Xc_S} \lVert \widehat{\psi}^t(x) - \psi^t(x) \rVert_2 \\
&\leq \max_{\substack{x\in\Xc_{S},y\in\Yc\\w\in\Wc}}  \left|(\widehat{V}(x)^{-1})_y\cdot w - (V^e(x)^{-1})_y\cdot w\right| + \eta\max_{x\in\Xc_S} \lVert \widehat{\psi}^t(x) - \psi^t(x) \rVert_2,\numberthis\label{equ::t1::C1}
\end{align*}
where $(a)$ follows from~\eqref{equ::t1::bnd_g} and~\eqref{equ::t1::bnd_rho}  in the proof of Proposition~\ref{pro_EY}. Briefly returning to the original notation, we can bound $\mathbf{C_1}+\mathbf{C_2}$ as
\begin{align}
&\widehat{\R}_{\widehat{S}}^t(h_{\widehat{C},\widehat{S}};\boldsymbol{\rho}^e_{\widehat{C},\widehat{S}},g_{\widehat{C},\widehat{S}}^t) - \widehat{\R}_{\widehat{S}}^t(h_{\widehat{C},\widehat{S}};\boldsymbol{\rho}^e_{\widehat{C},\widehat{S}},\psi_{\widehat{C},\widehat{S}}^t) \\
&\quad + \widehat{\R}_{S^*}^t(h_{C^*,S^*};\widehat{\boldsymbol{\rho}}_{C^*,S^*},\widehat{\psi}^{t}_{C^*,S^*}) - \widehat{\R}_{S^*}^t(h_{C^*,S^*};\boldsymbol{\rho}^e_{C^*,S^*},\psi^{t}_{C^*,S^*})\\
& \leq 2\max_{C,S}|\widehat{\R}_S^t(h_{C,S};\boldsymbol{\rho}^e_{C,S},\psi_{C,S}^t) - \widehat{\R}_S^t(h_{C,S};\widehat{\boldsymbol{\rho}}_{C,S},\widehat{\psi}_{C,S}^t)|.
\end{align}
Thus, using~\eqref{equ::t1::C1}, $\mathbf{C_1}+\mathbf{C_2}$ is bounded by 
\begin{align*}
    2&\max_{\substack{x\in\Xc_{S},y\in\Yc\\w\in\Wc,C,S}}  \left|(\widehat{V}_{C,S}(x)^{-1})_y\cdot w - (V^e_{C,S}(x)^{-1})_y\cdot w\right| + 2\eta\max_{\substack{x\in\Xc_{S}\\C,S}} \lVert \widehat{\psi}_{C,S}^t(x) - \psi^t_{C,S}(x) \rVert_2 \\
    &\leq \max_{\substack{x\in\Xc_{S},y\in\Yc\\C,S}}  2d\lVert(\widehat{V}_{C,S}(x)^{-1})_y - (V^e_{C,S}(x)^{-1})_y\rVert_2 + 2\eta\max_{x\in\Xc_S,C,S} \lVert \widehat{\psi}_{C,S}^t(x) - \psi_{C,S}^t(x) \rVert_2.  \numberthis\label{equ::t1::C}
\end{align*}

\noindent\textbf{Term $\mathbf{D}$}: As $(\widehat{C},\widehat{S})$ minimizes the objective in~\eqref{equ::opt_dimp_finite}, 
\begin{equation}
    \widehat{\R}_{\widehat{S}}^t(h;\widehat{\boldsymbol{\rho}},\widehat{\psi}^t) + \Lambda_{\widehat{C},\widehat{S}}  \leq \widehat{\R}_{S^*}^t(h^*;\widehat{\boldsymbol{\rho}}^*,\widehat{\psi}^{t,*})  + \Lambda_{C^*,S^*}.
\end{equation}
Thus, 
\begin{align*}
&\widehat{\R}_{\widehat{S}}^t(h;\widehat{\boldsymbol{\rho}},\widehat{\psi}^t) - \widehat{\R}_{S^*}^t(h^*;\widehat{\boldsymbol{\rho}}^*,\widehat{\psi}^{t,*}) \\
    &\quad\leq \Lambda_{C^*,S^*} - \Lambda_{\widehat{C},\widehat{S}} \\
    &\quad \leq \Lambda_{C^*,S^*}. \numberthis\label{equ::t_D}
\end{align*}

\noindent\textbf{Term $\mathbf{E}$}: 
Since $\rho^{e,*}_y$ and $\psi^{t,*}$ are EY-separable according to~\eqref{equ::opt_wdimp_finite_hat}, then  following~\eqref{equ::EY_to_R}, $\widetilde{\R}_{S^*}^t(h^*;\boldsymbol{\rho}_e^*,\psi^{t,*}) = R_{S^*}^t(h^*)$.

\medskip
Putting the terms together completes the proof Theorem~\ref{the:2}. 

\end{proof}

\end{document}